\documentclass[superscriptaddress,onecolumn,aps,prd,showpacs,floatfix,11pt,nofootinbib]{revtex4-1}
\usepackage{amssymb,amsmath}
\usepackage[dvips]{graphics,color}
\usepackage[colorlinks,hyperindex]{hyperref}
\usepackage{epsfig}\usepackage{float,stmaryrd,textcxomp,bm,mathbbol}\usepackage{float,xcolor,upgreek,yfonts,tikz}
\usepackage{graphicx}

\newcommand{\ba}{\begin{eqnarray}}
\newcommand{\ea}{\end{eqnarray}}
\newcommand{\enge}{\end{equation}}

\newcommand{\beq}{\begin{eqnarray}}
\newcommand{\benu}{\begin{enumerate}}
\newcommand{\enu}{\end{enumerate}}
\newcommand{\eeq}{\end{eqnarray}}

\hypersetup{hidelinks,backref=true,pagebackref=true,hyperindex=true,colorlinks=true,breaklinks=true,urlcolor= blue}
\hypersetup{%
  colorlinks = true,
  linkcolor  = blue,
    citecolor = cyan,
}
\newcommand\orcidroldao{{\href{https://orcid.org/0000-0003-3978-532X}{\orcidicon}}}
\newcommand{\orcidicon}{%
	\begin{tikzpicture}
	\draw[lime, fill=lime] (0,0)
		circle [radius=0.16]
		node[white] {{\fontfamily{qag}\selectfont \tiny ID}};
	\draw[white, fill=white] (-0.0625,0.095)
		circle [radius=0.007];
	\end{tikzpicture}	\hspace{-2mm}
}

\definecolor{purple}{rgb}{1,0,1}
\definecolor{lime}{HTML}{A6CE39} 

 \newcommand{\bea}{\begin{eqnarray}}
 \newcommand{\eea}{\end{eqnarray}}

\usepackage{hyperref}
\usepackage{xcolor}

 \newcommand{\bltx}{\textcolor{black}}

\begin{document}

\title{Deformations of the AdS-Schwarzschild black brane and the shear viscosity  
of the quark-gluon plasma 
}

\author{Roldao da Rocha\orcidroldao\!\!}
\email{roldao.rocha@ufabc.edu.br}
\affiliation{Federal University of ABC, Center of Mathematics,  Santo Andr\'e, 09210-580, Brazil.}


\begin{abstract}
Deformations of the AdS$_5$-Schwarzschild black brane, implemented in the AdS/CFT membrane paradigm, 
are scrutinized in the dual viscous hydrodynamic infrared limit. The latest experimental data analyses, regarding the shear viscosity-to-entropy density ratio of the quark-gluon plasma 
produced by heavy-ion collisions at the LHC and RHIC,  are shown to constrain these deformations severely. Although corroborating with the robustness of the standard AdS$_5$-Schwarzschild black brane against deformations, there is still a margin for mild deformations which may carry  2-loop quantum corrections to gravity, whose implications to the strongly-coupled dual field theory are addressed and discussed. \end{abstract}
\maketitle

\section{Introduction}

AdS/CFT provides robust tools to investigate   
  field theories in the strongly-coupled regime, whose  hydrodynamical infrared (IR) limit corresponds to long-length scales.  
 In the AdS bulk, weakly-coupled gravity is dual to the strongly-coupled conformal field theory (CFT) on the AdS boundary.  \cite{Maldacena:1997re}. Black brane solutions of type II supergravity and their near-horizon geometry are central objects of the AdS/CFT correspondence. At finite temperature, the AdS bulk geometry becomes an AdS-Schwarzschild black brane with an event horizon, corresponding to the low-energy limit\footnote{In this limit, the  AdS$_5\times S^5$ geometry is suppressed by the effective AdS$_5$ one.} of the metric equipping a stack of $N_c$ non-extremal  Dirichlet 3-branes ($D_3$-branes). The dual structure is an $\mathcal{N} = 4$ super-Yang--Mills theory with gauge symmetry SU($N_c$), with the beta-function equaling zero and energy-scale-independent coupling constant, at finite temperature equivalent to the Hawking temperature of the black brane, taking into account the 
large-$N_c$ limit, for large ’t Hooft coupling $\lambda = g_{\scalebox{.65}{\textsc{YM}}}^2N_c\to\infty$ \cite{Witten:1998qj,Gubser:1998bc}.  These limits are required to keep the
curvature corrections and string-loop corrections small enough in the calculations on the gravity side.
In the holographic duality dictionary, the long-length scale regime of the CFT living on the AdS boundary represents the near-horizon limit in the gravitational background. 
When both the ’t Hooft coupling and the set of $D_3$-branes are large, the curvature induced by the black brane geometry is tiny enough for classical gravity to describe the near-horizon geometry of the stack of $D_3$-branes. Hence, two equivalent versions of the same physical system emerge, where a strongly-coupled gauge theory residing on the stack of $D_3$-branes accounts for classical gravity on AdS$_5\times S^5$ \cite{Cremonini:2011iq}.

 Black holes and black branes can encode fluid flows on their event horizon in the membrane paradigm, whose low-energy regime corresponds to a strongly-coupled field theory.  
Solutions of Einstein's field equations in the AdS bulk are translated into relativistic viscous hydrodynamics regulated by Navier--Stokes equations on the AdS boundary, in the fluid/gravity duality  \cite{Bhattacharyya:2008jc,Bhattacharyya:2008kq,Haack:2008cp,Ferreira-Martins:2019svk,Pinzani-Fokeeva:2014cka,Rocha:2022ind,deBoer:2014xja}. 
Conformal symmetries on the gravity side of the duality are related to accelerated boost symmetries of Navier--Stokes equations, in the field-theoretical side, encompassed by the BDNK  (Bemfica--Disconzi--Noronha--Kovtun) setup  \cite{Bemfica:2020xym,Bemfica:2017wps,Kovtun:2019hdm,Hoult:2020eho}. In the membrane paradigm, hydrodynamical fluid flows reside on the AdS boundary at the so-called cutoff manifold, emulating the ultraviolet (UV) cutoff in the dual field theory \cite{20}. Black holes play the role of dissipative branes at finite temperature, with entropy, surface viscosity, and electrical resistivity 
 \cite{Kovtun:2003wp,Parikh:1997ma,Price:1986yy,hub,deBoer:2015ija}. 
The membrane paradigm makes the bulk black hole geometry match the codimension-one viscous fluid that emerges at the stretched horizon, which represents a timelike manifold slightly outside the black hole event horizon, capable of thermalizing, absorbing, and radiating information. Microstates describing the black hole represent dynamical degrees of freedom
related to the physical membrane infinitesimally near the black hole event horizon. It mimics surrogates for any global event horizon in phenomenological prescriptions for black
hole evolution \cite{Susskind:1993if}. In the membrane paradigm, 
the relaxation time for the shear viscous stress can be evaluated for AdS-Schwarzschild black branes and $D_p$-branes as well \cite{Natsuume:2007ty}. 
The AdS boundary mirrors a 4-dimensional brane, in braneworld models. Branes have (hyper)surface tension, which is fine-tuned to both the brane and the bulk cosmological 
constants \cite{rs1,daRocha:2017cxu,Bazeia:2024gvy}. 

General relativity (GR) describes a theory of gravity in an ideally rigid brane presenting infinite tension. Nevertheless, the most strict phenomenological bound for the finite value of the brane tension was obtained by the analysis of the cosmic microwave background (CMB)  anisotropy by the Wilkinson Microwave Anisotropy Probe (WMAP) in Ref. \cite{Fernandes-Silva:2019fez}.  This bound 
establishes a reliable low-energy limit permitting AdS/CFT membrane paradigm analogs for classical solutions in GR  \cite{Iqbal:2008by,mgd2,ssm1,Shiromizu:2001jm}. Einstein's field equations near the horizon of the black hole reduce to the Navier--Stokes equations for the fluid flow \cite{Rangamani:2009xk}. Soft-hair excitations in the  gravitational background are dual to soft hair in the Navier--Stokes equations, represented by an effective term coupling the Ricci tensor to the fluid flow velocity, which corrects the Laplacian in the diffusion term carrying the kinematic viscosity \cite{Ferreira-Martins:2021cga}. A fluid at the black hole horizon mimics a fluid at the AdS boundary introducing a convenient dictionary, associating braneworld models to the AdS/CFT  membrane paradigm \cite{Eling:2009sj,Antoniadis:1998ig,Antoniadis:1990ew,daRocha:2012pt,Abdalla:2009pg}.

Relativistic hydrodynamics plays a prominent role in phenomenological and theoretical approaches of field theory. In particular,  experiments at the Relativistic Heavy Ion Collider (RHIC) corroborate with computational simulations regarding relativistic hydrodynamics \cite{Baier:2007ix,Torrieri:2020ezm,Karapetyan:2018yhm,Karapetyan:2018oye,Karapetyan:2021vyh}. 
Fluid/gravity duality 
is an efficient setup to 
compute transport and response coefficients associated with  black branes, using the hydrodynamical description of the dual field theory. This framework can be implemented when isometries are employed to perturb the horizon of black brane solutions of Einstein’s field equations in AdS. 
One can use the fluid/gravity duality to compute the shear viscosity-to-entropy density ratio ($\eta/s$), among other relevant transport and response coefficients.   The Kovtun--Son--Starinets (KSS) viscosity-bound saturation $\eta/s=1/4\pi$ was calculated for the $\mathcal{N} = 4$ SU($N_c$) super-Yang--Mills plasma, in the planar limit and for infinitely large ’t Hooft coupling \cite{Policastro:2001yc}, using the Kubo  formula to relate $\eta/s$ to the classical absorption cross-section of gravitons by black branes \cite{kss,Gubser:1997yh,BoschiFilho:2004ci}.

Weakly-coupled gravity has been employed as a dual scenario to explore strongly-coupled field theories, bringing robust advances for formulating correlators and the finite-temperature behavior at strong-coupled regimes  \cite{Aharony:1999ti}.  The hydrodynamical limit of the gauge/gravity duality is fruitful for investigating the quark-gluon
plasma (QGP), consisting of a transient QCD phase of deconfined ultra-dense and hot nuclear matter, which is completely locally thermalized, corroborating to the analysis of experimental data at the LHC and RHIC \cite{JETSCAPE:2020shq,Parkkila:2021yha,Bernhard:2019bmu,Nijs:2022rme,Nijs:2020ors,Yang:2022ixy,Rougemont:2023gfz,Dash:2019zwq,Finazzo:2016psx,Noronha-Hostler:2008kkf,Yang:2022yxa,Finazzo:2014cna,Denicol:2014xca,Hippert:2023bel,ALICE:2011ab,ALICE:2021adw,ALICE:2021klf}. 
As large-$N_c$ gauge theories in the AdS/CFT setup can 
reasonably describe QCD, the KSS viscosity-bound saturation can be applied to the QGP. Experiments in the RHIC and LHC have shown that the QGP behaves like a viscous fluid with tiny viscosity, being  the closest in nature to the KSS viscosity-bound saturation. It gets rid of using perturbative QCD in the study of the QGP and, hence, heavy-ion collisions can be studied by relativistic viscous hydrodynamics \cite{Song:2012ua}. One can thus probe the QGP through AdS/CFT and generalizations \cite{qgpexp2}. 
Evaluating the Kubo  formula in QCD determines the viscosity microscopically. Although it is an intricate task to be accomplished analytically at strong-coupling, lattice simulations become possible \cite{Nakamura:2004sy,Cucchieri:2007rg}. 
The QGP was also investigated close to the QCD crossover transition, in the temperature range $T \sim 150$ - $350$ MeV, where QCD
matter acts as a strongly-interacting, non-conformal, 
nearly perfect fluid. The gauge/gravity holographic
correspondence was employed to address strongly-coupled nonequilibrium phenomena in an expanding Universe at temperatures near the QCD phase transition \cite{Sanches:2014gfa}, which are currently beyond the scope of other non-perturbative approaches such as lattice gauge theory \cite{Heinz:2013th,Oliveira:2018ukh,DerradideSouza:2015kpt}.

Transport coefficients of the QGP, including the bulk and shear viscosities, the thermal and baryon conductivities, the jet quenching parameter, and diffusion coefficients were computed in Ref. \cite{Grefa:2022sav}, using Einstein--Hilbert gravity coupled to a Maxwell gauge field and a dilaton. 
Ref. \cite{Niemi:2011ix} showed that the elliptic flow in heavy-ion collisions at RHIC is dominated
by the shear viscosity in the hadronic phase and the phase transition region.
 Relativistic dissipative fluid dynamics have been thoroughly scrutinized, with the calculation of transport and response coefficients for the non-conformal QGP from holography \cite{Hatta:2014jva,Rocha:2023ilf,Brito:2021iqr,Diles:2019uft}. Also, Ref. \cite{Ghosh:2015mda} computed the shear and bulk viscosities using the quark
thermal width, whereas Ref. \cite{Gardim:2014tya} calculated these transport coefficients for anisotropic flows.
The anisotropic shear viscosity can be addressed with anisotropic black branes \cite{Brito:2019ose}.  
 Transport coefficients were computed for the hot QGP at finite chemical potential \cite{Soloveva:2019xph}. 
Many other seminal aspects of the QGP emulating a quantum field theory that is dual to holographic black holes, and subsequent calculation of the bulk and shear viscosities were explored in Refs. 
\cite{Grefa:2021qvt,Karapetyan:2023sfo,Karapetyan:2020yhs,Karapetyan:2016fai,Critelli:2017oub,Rougemont:2015ona,Rougemont:2015wca,MartinContreras:2021bis}. Models for rotating QGP were scrutinized in Ref. \cite{Braga:2022yfe,Braga:2023qee,Braga:2023qej}. 
Phenomenological aspects of AdS$_5$-Schwarzschild black branes deformations were explored in the context of phase transitions in AdS/QCD and the holographic entanglement entropy  
\cite{daRocha:2021xwq}. 
AdS$_4$ generalized extremal branes were scrutinized in the context of the membrane paradigm of AdS/CFT, whose near-horizon methods made it possible to calculate the shear viscosity-to-entropy density ratio. Holographic superconductors and their coefficients of response and transport in the dual condensed matter theory probed  experimental features  for deformations of the 
AdS$_4$-Reissner-Nordström black brane \cite{Ferreira-Martins:2019wym,daRocha:2023waq,Meert:2024dud}. 
 Also, 1-loop quantum corrections to $\eta/s$ were addressed consistently, complying with phenomenological aspects underlying the QGP \cite{Kuntz:2022kcw}.  
Effective holographic models for QCD were also considered by deformations of the AdS-Schwarzschild black brane, whose viscosity coefficients were computed and analyzed \cite{Ballon-Bayona:2021tzw}.

The 3-brane solution to the equations of type IIB supergravity was originally derived in Ref.  \cite{Horowitz:1991cd}, comprising the Ramond--Ramond self-dual 5-form field strength coupling to the 3-brane and a dilaton field, with blackening factor given by the AdS-Schwarzschild one \cite{Gubser:1996de}. 
The AdS$_5$-Schwarzschild black brane can be deformed by using the embedding algorithm and using the Hamiltonian and momentum constraints from the Arnowitt--Deser--Misner (ADM) formalism for static configurations of the metric field \cite{Casadio:2003jc}. Black brane embeddings were also  implemented for the case of anisotropic branes \cite{Avila:2016mno}. It permits a family of deformations of the AdS$_5$-Schwarzschild black brane \cite{Ferreira-Martins:2019svk,daRocha:2021xwq}. Here we derive deformed AdS$_5$-Schwarzschild  black brane solutions and use the shear viscosity-to-entropy density ratio, ${\eta}/{s}$, and the  black brane temperature, to impose viscosity bounds to the free parameter governing the family of new black brane solutions.
The latest experimental data for the ratios  $\eta/s$   of the QGP, produced by heavy-ion collisions at LHC and RHIC, will be utilized to bound the parameter regulating deformations of the AdS$_5$-Schwarzschild black brane. 
We will show that mild deformations of the AdS$_5$-Schwarzschild black brane might allow and encode higher-order curvature terms driving 2-loop quantum corrections to gravity. We aim to provide robust evidence on how  deformations of AdS$_5$-Schwarzschild black branes, on the gravity side, are constrained by thermodynamical and hydrodynamical features of the dual field theory, and the implications thereof. The analysis of experimental data about the QGP at LHC and RHIC will be shown to support a narrower range for deformations of the AdS$_5$-Schwarzschild black brane. 
This work is organized as follows: in Sec. \ref{sec:lrt}, hydrodynamics and the linear response theory are introduced in AdS/CFT, where the Kubo  formula for the shear viscosity is presented. Sec.  \ref{sec:eta_s_sads_5} is devoted to constructing the AdS$_5$-Schwarzschild black brane deformations. The solution of the Einstein--Hilbert action also contains a counterterm to avoid divergences, and the Gibbons--Hawking term yields the partition function for the dual theory, from the GKPW relation. The free energy, the entropy,  the pressure, and the specific heat are computed as state functions in the canonical ensemble. In Sec. \ref{sec:eta_s_ads_mgd_5}, the $\eta/s$ ratio, the black brane temperature, and the shear mode damping constant are addressed to the deformed AdS$_5$-Schwarzschild black brane. 
Sec. \ref{rhic} shows how to constrain the parameter responsible for deforming the black brane from experimental data of the QGP and heavy-ion collisions 
at LHC and RHIC. The JETSCAPE Bayesian model  \cite{JETSCAPE:2020shq},  the results by the Duke group \cite{Bernhard:2019bmu}, the ones by the  Jyv\"askyl\"a-Helsinki-Munich group \cite{Parkkila:2021yha}, and the results by the MIT-Utrecht-Gen\`eve group  \cite{Nijs:2022rme,Nijs:2022rme} are explored in detail to bound the black brane deformation parameter and to show that the family of deformed AdS$_5$-Schwarzschild black branes is robust under deformations, however mild deformations are allowed, alternatively carrying higher-order curvature terms driving 2-loop quantum corrections to gravity.
 The concluding remarks are presented in Sec. \ref{sec:5}. Appendix \ref{app2} contains the deformation procedure setup obtained by the embedding algorithm, with the Hamiltonian and momentum constraints. Appendix \ref{app3} shows the deformed AdS$_5$-Schwarzschild  black brane as an exact solution of higher-order curvature terms. Despite addressing the robustness of the AdS$_5$-Schwarzschild black brane against deformations, a strict margin for mild deformations carrying higher-order curvature terms driving quantum corrections to gravity can still be used to analyze and study the QGP and other physical systems in QCD. Hence the question of determining the effect of the higher-curvature and higher-derivative corrections to the thermal and hydrodynamical  properties of the dual gauge theory reduces to the ordinary problem of investigating the properties of deformed AdS$_5$-Schwarzschild  black branes within a certain 5-dimensional gravity action with a universal set of higher-derivative corrections involving cubic terms in the curvature and the combination of quadratic terms in the covariant derivatives of the Riemann and Ricci tensors, and the scalar curvature.

\section{Hydrodynamics, linear response theory, and the Kubo  formula}
\label{sec:lrt}

\par 
In AdS/CFT, fluid dynamics comprises the low-energy effective description of any interacting QFT. The so-called hydrodynamic limit is characterized by the long-wavelength, low-energy regime, and it is often applicable to describe conserved quantities \cite{Rangamani:2009xk}. 
Hydrodynamics does not incorporate all the features of a microscopic theory. Transport and response coefficients can be determined from experimental data or computed from the 
microscopic theory. Kubo relations provide the correspondence among transport coefficients and correlation
functions in QFT. The approach to black hole gravitational perturbations indicates that fluctuations of the stretched horizon have to encode the complete collection of
hydrodynamical modes. Methods in fluid/gravity correspondence must replicate wave modes describing the sound propagation and non-propagating modes, including shear and diffusion. Hydrodynamics, as an effective theory, designates the dynamics ruling any thermal system whose time and length scales are large when contrasted to microscopic scales. Hydrodynamics proposes observing a shear diffusion mode whose diffusion constant is proportional to the shear viscosity, which measures the amount of dissipation in a simple fluid.

\par The macroscopic variables encoded in the energy-momentum stress tensor, $T^{\mu \nu}$, along with its conservation law, $\nabla_\mu T^{\mu \nu} = 0$, describe a simple fluid. At first order, it reads  \cite{Rangamani:2009xk,Bhattacharyya:2008jc}
\begin{equation} \label{TEM:diss}
T^{\mu \nu} = p\left(g^{\mu \nu} + u^\mu u^\nu  \right ) + \rho u^\mu u^\nu  + \tau^{\mu \nu} \ ,
\end{equation}
where $\tau^{\mu \nu}$ carries dissipative effects, whereas the normalized fluid velocity field is denoted by $u^\mu(x^\nu)$, its pressure field by $p(x^\mu)$, and its rest-frame energy density by $\rho(x^\mu)$. Eq. \eqref{TEM:diss} yields both the continuity and the Navier--Stokes equations.  For any theory described by a general action $S$, the coupling of an operator ${\scalebox{.93}{$\mathcal{O}$}}$ to an external source ${\scalebox{.98}{$\upvarphi$}}^{\scalebox{.65}{$(0)$}}$ can be implemented by the mapping \cite{kss,kss}
\begin{equation}
	S\mapsto \mathring{S} = S + \int {\scalebox{.95}{$\mathrm{d}$}}^4x\, {\scalebox{.98}{$\upvarphi$}}^{\scalebox{.65}{$(0)$}}(t, \vec{x})\, {\scalebox{.93}{$\mathcal{O}$}}(t, \vec{x})\ . 
\end{equation}
The linear response theory takes into account the response of ${\scalebox{.93}{$\mathcal{O}$}}$ at first order in the source ${\scalebox{.98}{$\upvarphi$}}^{\scalebox{.65}{$(0)$}}$. 
Dynamical features of a thermal gauge theory can be read from its Green functions which, in AdS/CFT, can be computed using the dual weakly-coupled gravity \cite{Son:2002sd}. 
The 1-point function reads 
\begin{equation}
\updelta \left \langle {\scalebox{.93}{$\mathcal{O}$}} (\omega, \bm{\mathfrak{q}})\right \rangle = - G_{\scalebox{.65}{$R$}}^{{\scalebox{.63}{$\mathcal{O}$}}, {\scalebox{.63}{$\mathcal{O}$}}} (\omega, \bm{\mathfrak{q}}) {\scalebox{.98}{$\upvarphi$}}^{\scalebox{.65}{$(0)$}}(\omega, \bm{\mathfrak{q}}) \ ,
\label{eq:response_def}
\end{equation}
\noindent where $G_{\scalebox{.65}{$R$}}^{{\scalebox{.63}{$\mathcal{O}$}}, {\scalebox{.63}{$\mathcal{O}$}}} (\omega, \bm{\mathfrak{q}})$ is the retarded Green's function \cite{Natsuume:2014sfa}. 
The shear viscosity comes upon the response of the fluid, which is the field-theoretical dual to the AdS-Schwarzschild black brane solution,  undergoing thermal and viscous internal forces. By coupling gravitational perturbations to the hydrodynamical fluid flow, and measuring the response of the energy-momentum tensor under gravitational perturbations, the Kubo  formula establishes the shear viscosity as a transport coefficient associated with the  retarded Green's function \cite{Natsuume:2014sfa,Ghosh:2019ubc}. 
The response of $\tau^{\mu \nu}$ under gravitational fluctuations can be implemented by an arbitrary off-diagonal perturbation $h_{xy}^{\scalebox{.65}{$(0)$}}$, leading to the perturbed metric \cite{Bhattacharyya:2008jc}:
\begin{equation}
g_{\mu \nu}^{\scalebox{.65}{$(0)$}}{\scalebox{.95}{$\mathrm{d}$}}x^\mu {\scalebox{.95}{$\mathrm{d}$}}x^\nu = \eta_{\mu \nu} {\scalebox{.95}{$\mathrm{d}$}}x^\mu {\scalebox{.95}{$\mathrm{d}$}}x^\nu + 2h_{xy}^{\scalebox{.65}{$(0)$}}(t) {\scalebox{.95}{$\mathrm{d}$}}x {\scalebox{.95}{$\mathrm{d}$}}y.
\label{eq:perturbed_metric}
\end{equation} 
In the fluid/gravity correspondence, long-wavelength perturbations on top of an equilibrium state in the hydrodynamic fluid flow  correspond to long-wavelength ripples on the deformed AdS$_5$-Schwarzschild  black brane. The correlation functions can be computed using linear response theory, in the dual field theory. The response is
governed by the retarded correlation function. In the case of shear viscosity, the correlation function can be obtained from  \cite{kss}
\begin{equation}
\lim_{\bm{\mathfrak{q}} \to \bm{0}}\updelta \left \langle \tau^{xy} (\omega, \bm{\mathfrak{q}}) \right \rangle = i \omega \eta h_{xy}^{\scalebox{.65}{$(0)$}}=- G_{\scalebox{.65}{$R$}}^{xy, xy} h_{xy}^{\scalebox{.65}{$(0)$}}\,,
 \label{oooo}
\end{equation}
where 
\beq
G_{\scalebox{.65}{$R$}}^{xy, xy}= -i \int^{\infty}_{-\infty} {\rm d}^4x\, e^{i\omega t -i\vec{q}\cdot\vec{x}}
\theta(t)  \left\langle [T^{xy}(t, \vec{x} ), T^{xy}(0,\vec{0}) ] \right\rangle.\eeq
Matching
the correlation function from linear response theory to the hydrodynamic correlator gives the Kubo formula for the shear viscosity \cite{Schafer:2009dj}:
\begin{equation}
\eta = - \lim_{\substack{{\omega \rightarrow 0}\\\bm{\mathfrak{q}} \to \bm{0}}} \frac{1}{\omega} \Im\left (G_{\scalebox{.65}{$R$}}^{xy, xy} (\omega, \bm{\mathfrak{q}}) \right ). \label{xxx}
\end{equation}
The imaginary part of the retarded correlator is a measure
of dissipation. 
Both the small-frequency and zero-momentum limits in Eq. (\ref{xxx}) encode large-distance and long-timescale regimes. Transport coefficients, in particular the shear viscosity, 
are parameters in effective low-energy hydrodynamics, driving the macroscopic responses of the fluid flow.

 Computation of the retarded Green's function is straightforwardly achieved, once the Gubser--Klebanov--Polyakov--Witten (GKPW) relation, which states that the partition function of the CFT on the boundary equals the partition function of the dual gravitational theory,  is regarded \cite{Witten:1998qj,Gubser:1996de,Gubser:1998bc}.
It yields the following expression for the 1-point function,
\begin{equation}
\left \langle {\scalebox{.93}{$\mathcal{O}$}} \right \rangle_S = \frac{\updelta \mathring{S}[{\scalebox{.98}{$\upvarphi$}}^{\scalebox{.65}{$(0)$}}]}{\updelta {\scalebox{.98}{$\upvarphi$}}^{\scalebox{.65}{$(0)$}}} \ .
\label{eq:one_point}
\end{equation}
\noindent One considers AdS$_5$ bulk gravity governed by GR, with negative cosmological constant $\Lambda_5$. Therefore the action reads 
\begin{equation}\label{eq:fullAction}
	S=\frac{1}{16\pi}\int {\scalebox{.95}{$\mathrm{d}$}}^{5}x\sqrt{-g}\left(R-2\Lambda_5\right)+S_{\textsc{mat}} \ ,
\end{equation}
 \noindent where $S_{\textsc{mat}}$ is specified by the boundary theory of interest. For example, the matter action for a massless scalar field corresponds to a kinetic term. Without taking into account $S_{\textsc{mat}}$, a solution to the equations of motion coming from Eq. \eqref{eq:fullAction} is the  AdS$_5$-Schwarzschild black brane,
\begin{equation}
{\scalebox{.95}{$\mathrm{d}$}}s^2 = -\frac{r_{\scalebox{.65}{$0$}}^2}{u^2} f(u) {\scalebox{.95}{$\mathrm{d}$}}t^2 + \frac{1}{u^2 f(u)} {\scalebox{.95}{$\mathrm{d}$}}u^2 + \frac{r_{\scalebox{.65}{$0$}}^2}{u^2} \delta_{ij} {\scalebox{.95}{$\mathrm{d}$}}x^i {\scalebox{.95}{$\mathrm{d}$}}x^j \ ,
\label{eq:sads_5_u}
\end{equation}
\noindent  with $u=r_{\scalebox{.65}{$0$}}/r$, and $r_{\scalebox{.65}{$0$}}$ is the horizon radius, for where $f(u) = 1-u^4$. Hence $u=1$ locates the horizon, whereas $u=0$ is the black brane boundary corresponding to $r\to\infty$.  For $u\rightarrow 0$, Eq. \eqref{eq:sads_5_u} reads
\begin{eqnarray}
{\scalebox{.95}{$\mathrm{d}$}}s^2= \frac{r_{\scalebox{.65}{$0$}}^2}{u^2} \left (-{\scalebox{.95}{$\mathrm{d}$}}t^2 + \frac{1}{r_{\scalebox{.65}{$0$}}^2}{\scalebox{.95}{$\mathrm{d}$}}u^2 + \delta_{ij} {\scalebox{.95}{$\mathrm{d}$}}x^i {\scalebox{.95}{$\mathrm{d}$}}x^j \right ) \ .
\label{eq:asym_ads}
\end{eqnarray}
The 1-point function \eqref{eq:one_point} is dependent  exclusively  on $S_{\textsc{mat}}$. Assuming ${\scalebox{.98}{$\upvarphi$}} = {\scalebox{.98}{$\upvarphi$}}(u)$, the action governing a massless scalar field on the boundary wits
\begin{eqnarray}
\!\!\!\!\!\!\!\!\!\!\!S_{\textsc{mat}} \sim \lim_{u\to0}\int{\scalebox{.95}{$\mathrm{d}$}}^4x  \frac{r_{\scalebox{.65}{$0$}}^4}{2u^3} {\scalebox{.98}{$\upvarphi$}} \dot{\scalebox{.98}{$\upvarphi$}} + \int {\scalebox{.95}{$\mathrm{d}$}}^5x 
\frac{r_{\scalebox{.65}{$0$}}^4}{2u^3}\left ( \ddot{\scalebox{.98}{$\upvarphi$}} \!-\!\frac{3}{u} \dot{\scalebox{.98}{$\upvarphi$}}\right )\! {\scalebox{.98}{$\upvarphi$}}.
\label{eq:onshell_quase}
\end{eqnarray}
The (on-shell) action leads to a surface term on the AdS boundary, making  
Eq. (\ref{eq:onshell_quase}) to yield the equation of motion for the scalar field with asymptotic solution $
 {\scalebox{.98}{$\upvarphi$}} \sim {\scalebox{.98}{$\upvarphi$}}^{\scalebox{.65}{$(0)$}} \left (1 + {\scalebox{.98}{$\upvarphi$}}^{\scalebox{.65}{$(1)$}} u^4 \right )$. 
Superseding it into Eq. \eqref{eq:onshell_quase} yields 
\begin{equation}\label{eq:responsegeneralgkpw}
	\left \langle {\scalebox{.93}{$\mathcal{O}$}} \right \rangle_S = 4r_{\scalebox{.65}{$0$}}^4 {\scalebox{.98}{$\upvarphi$}}^{\scalebox{.65}{$(1)$}} {\scalebox{.98}{$\upvarphi$}}^{\scalebox{.65}{$(0)$}} = \updelta \left \langle {\scalebox{.93}{$\mathcal{O}$}}\right\rangle.
\end{equation}
The retarded Green's function is therefore obtained when Eq. \eqref{eq:responsegeneralgkpw} is compared 
 to Eq. \eqref{eq:response_def}, as \begin{equation}
 \lim_{\bm{\mathfrak{q}} \to \bm{0}}G_{\scalebox{.65}{$R$}}^{{\scalebox{.63}{$\mathcal{O}$}}, {\scalebox{.63}{$\mathcal{O}$}}} (\bm{\mathfrak{q}}) = -4 r_{\scalebox{.65}{$0$}}^4 {\scalebox{.98}{$\upvarphi$}}^{\scalebox{.65}{$(1)$}}.
\end{equation}
\noindent

\section{The deformed AdS$_5$-Schwarzschild black brane, GKPW relation, and thermodynamic functions}
\label{sec:eta_s_sads_5}

\par The 5-dimensional Einstein--Hilbert action with a negative cosmological constant has the AdS$_5$-Schwarzschild black brane as a solution. 
Black branes with large radii
can be approximated by a planar black hole with a
horizon invariant by translations \cite{Witten:1998zw,Bilic:2022psx}.
Ref. \cite{Witten:1998zw} considered a maximally supersymmetric Yang--Mills theory on the $p + 1$ dimensional target space world-volume of $N_c$ $D_p$-branes. When the supersymmetric theory is compactified on a circle, fermions achieve mass, which is inversely proportional to the circle radius. In contrast, bosons acquire their masses from loop diagrams at energies below the one that makes them decouple. Pure QCD in $p$ dimensions corresponds to the effective field theory at large distances compared to the circle radius. Finite temperature sets in, considering the circle radius inversely proportional to the temperature  
\cite{Witten:1998zw,Aharony:1999ti}. For the particular case where $p=3$, the zero-temperature solution of the $D_3$-brane corresponds to the $\mathcal{N} = 4$ super-Yang--Mills theory at zero temperature describing fluctuations of the branes \cite{Tong:2005nf}. A 
stack of $N_c$ $D_3$-branes has the metric 
\beq\label{t0ads}
{\scalebox{.95}{$\mathrm{d}$}}s^2 = \left(1+\frac{L^4}{r^4}\right)^{-1/2}(-{\scalebox{.95}{$\mathrm{d}$}} t^2 + \delta_{ij}{\scalebox{.95}{$\mathrm{d}$}}x^i {\scalebox{.95}{$\mathrm{d}$}}x^j)+\left(1+\frac{L^4}{r^4}\right)^{1/2}({\scalebox{.95}{$\mathrm{d}$}}r^2+r^2 {\scalebox{.95}{$\mathrm{d}$}}\Omega_5^2),\eeq
where $L^4 =\frac{N_c}{2\pi^2 T_3}$,   $T_3\sim N_c/(g_s\ell_s^4)$ is the $D_3$-brane tension,  ${\scalebox{.95}{$\mathrm{d}$}}\Omega_5^2$ is the 5-dimensional solid angle,   $g_s\sim g_{\scalebox{.65}{\textsc{YM}}}^2$ denotes the string coupling constant for the emission and absorption of a closed string, and $\ell_s$ is the fundamental length scale of the string. 
The blackening factor in the metric (\ref{t0ads})
reflects the Newtonian potential $\phi_{\textsc{Newt}}=g_sN_c\ell_s^4/r^4$ obtained in the analysis regarding the way the $D_3$-brane curves spacetime \cite{Natsuume:2014sfa}.
The near-horizon limit ${r\ll R}$ implies the Taylor approximation $\left(1+\frac{L^4}{r^4}\right)^{-1/2}\approx \frac{L^2}{r^2}$, yielding the metric \eqref{t0ads} to be essentially the ${\rm AdS}_5\times S^5$ metric: \beq\label{ads5s5}
{\scalebox{.95}{$\mathrm{d}$}}s^2 = \frac{r^2}{L^2}(-{\scalebox{.95}{$\mathrm{d}$}}t^2 + \delta_{ij}{\scalebox{.95}{$\mathrm{d}$}}x^i {\scalebox{.95}{$\mathrm{d}$}}x^j)+\frac{L^2}{r^2}{\scalebox{.95}{$\mathrm{d}$}}r^2+{L^2} {\scalebox{.95}{$\mathrm{d}$}}\Omega_5^2.\eeq 
At this point, $L$ can also be interpreted as the AdS radius of curvature at leading order.
The temporal component of the metric \eqref{ads5s5} carries the factor $\frac{r}{R}$ in ${\scalebox{.95}{$\mathrm{d}$}}t$ and, since $E=i\hbar \frac{\partial}{\partial t},$ then it implies that the additional dimension in AdS$_5$ corresponds to the 4-dimensional energy scale, which is useful in the AdS/QCD correspondence \cite{Araujo:2014kda}. More precisely, using Poincar\'e coordinates, $z\equiv L^2/r$, 
the ${\rm AdS}_5\times S^5$ metric \eqref{ads5s5} can be expressed as 
${\scalebox{.95}{$\mathrm{d}$}}s^2 = \frac{L^2}{z^2}(-{\scalebox{.95}{$\mathrm{d}$}}t^2 + \delta_{ij}{\scalebox{.95}{$\mathrm{d}$}}x^i {\scalebox{.95}{$\mathrm{d}$}}x^j+{\scalebox{.95}{$\mathrm{d}$}}z^2)+L^2 {\scalebox{.95}{$\mathrm{d}$}}\Omega_5^2,$ implying that in the $z\to0$ limit the original AdS$_5$ metric  ${\scalebox{.95}{$\mathrm{d}$}}s^2 = \frac{L^2}{z^2}(-{\scalebox{.95}{$\mathrm{d}$}}t^2 + \delta_{ij}{\scalebox{.95}{$\mathrm{d}$}}x^i {\scalebox{.95}{$\mathrm{d}$}}x^j)$ is recovered, with boundary given by the 4-dimensional Minkowski spacetime.

In the context of holography, the UV limit of the field theory lives on the AdS boundary. Asymptotically AdS spaces are relevant, contributing to the partition function calculation. 
Non-extremal black branes are intrinsically dual to finite-temperature field theories and have metric given by  \cite{Horowitz:1991cd}
\beq\label{stack}
{\scalebox{.95}{$\mathrm{d}$}}s^2 =\left(1+\frac{L^4}{r^4}\right)^{-1/2}\left[-f(r){\scalebox{.95}{$\mathrm{d}$}}t^2 +\delta_{ij}{\scalebox{.95}{$\mathrm{d}$}}x^i{\scalebox{.95}{$\mathrm{d}$}}x^j\right] +\left(1+\frac{L^4}{r^4}\right)^{1/2}\left[f^{-1}(r) {\scalebox{.95}{$\mathrm{d}$}}r^2 +r^2{\scalebox{.95}{$\mathrm{d}$}}\Omega_5^2\right],\eeq
where $i,j = 1,2,3$, and
\beq
f(r) = 1 -\frac{r_{\scalebox{.65}{$0$}}^4}{r^4}.
\eeq
The black brane horizon is located at $r = r_{\scalebox{.65}{$0$}}$. 

In suitable limits, both the closed- and the open-string field theories comply with a decoupled flat-space closed-string sector multiplied by an equivalent sector. 
It implies that the open-string sector of the $\mathcal{N} = 4$ super-Yang--Mills theory is dual to a closed-string theory describing the graviton, inhabiting the near-horizon limit AdS$_5 \times S^5$ \cite{Lambert:2003zr}.
The near-extremal $D_3$-brane system is dual
to finite-temperature $\mathcal{N} = 4$ supersymmetric SU($N_c$) Yang--Mills theory, in the limit of large
$N_c$ and large ’t Hooft coupling \cite{Maldacena:1997re}. 
The near-horizon decoupling limit
makes the metric (\ref{stack}) to yield 
\beq\label{ads5s51}
{\scalebox{.95}{$\mathrm{d}$}}s^2 =-\frac{r^2}{L^2}\left[f(r){\scalebox{.95}{$\mathrm{d}$}}t^2 +\delta_{ij}{\scalebox{.95}{$\mathrm{d}$}}x^i{\scalebox{.95}{$\mathrm{d}$}}x^j\right] +\frac{L^2}{r^2}f^{-1}(r) {\scalebox{.95}{$\mathrm{d}$}}r^2 +L^2{\scalebox{.95}{$\mathrm{d}$}}\Omega_5^2.
\eeq
The first part is the AdS$_5$-Schwarzschild black brane  metric \eqref{ads5s51} in Poincaré coordinates, with planar horizon $\mathbb{R}^3$, whereas the second piece is the 5-sphere. The temporal component of the metric \eqref{ads5s51} contains the factor $r_{\scalebox{.65}{$0$}}^4/(r^2L^2)$, emulating the 5-dimensional Newtonian potential. 
The first part of the metric \eqref{ads5s51} is asymptotically AdS in the Poincar\'e patch, where the horizon is parallel to the Poincar\'e horizon,
 and it has inverse temperature $\beta= \pi L^2/r_{\scalebox{.65}{$0$}}$ 
\cite{Horowitz:1991cd}. 
The $r$ coordinate along the bulk again plays
the role of the energy scale in the Yang--Mills gauge theory, having the boundary at $r\to\infty$
corresponding to the UV in the dual gauge theory. The 5-sphere describes internal degrees of freedom associated with the $R$-symmetry group SU(4), intrinsic to theories carrying $\mathcal{N} = 4$ supersymmetries. For the metric \eqref{ads5s51}, the  AdS$_5$-Schwarzschild metric term 
with translationally invariant horizon 
obeys Einstein’s equations $R_{\mu\nu}= \frac{2\Lambda_5}3 g_{\mu\nu}$ in a 5-dimensional space with
cosmological constant $\Lambda_5 = -6/L^2$. 
At low-energy regimes, the effective geometry is the AdS$_5$-Schwarzschild one at finite temperature.

Tidal deformations of the AdS$_5$-Schwarzschild black brane metric can be implemented when one embeds the black brane onto a codimension-one AdS bulk. It emulates seminal black brane embedding approaches \cite{ssm1,Kanti:2002fx,Emparan:2017qxd,Alencar:2021ljc}, now applied for the AdS$_5$-Schwarzschild black brane. 
This procedure is detailed in the Appendix  \ref{app2}. One supposes the most general static and spherically symmetric black brane metric,  setting the AdS radius to unity, 
\begin{eqnarray}
\!\!\!\!\!\!\!\!\!\!{\scalebox{.95}{$\mathrm{d}$}}s^2 = -{r^2} F(r) {\scalebox{.95}{$\mathrm{d}$}}t^2 + \frac{1}{r^2 G(r)} {\scalebox{.95}{$\mathrm{d}$}}r^2 + {r^2} \delta_{ij} {\scalebox{.95}{$\mathrm{d}$}}x^i {\scalebox{.95}{$\mathrm{d}$}}x^j. \label{1a}
\end{eqnarray}
\noindent 
Metrics with $F(r) = G(r)$ have the Ricci tensor and the corresponding energy-momentum tensor to solve Einstein’s
equation with vanishing null-null components. It means that the restriction
of the Ricci tensor to the $t-r$ manifold is proportional to the metric itself. This condition is valid whenever the areal radius coordinate represents an affine parameter on the
radial null geodesics \cite{Jacobson:2007tj}. Here we want to keep the metric \eqref{1a} as general as possible to get deformations of the AdS$_5$-Schwarzschild black brane. 
It is worth emphasizing that when $F(r)=G(r)=f(r)=1-r_{\scalebox{.65}{$0$}}^4/r^4$ one recovers the usual AdS$_5$-Schwarzschild black brane metric \eqref{ads5s51}, 
whose asymptotic limit $r\to\infty$ is led to the AdS$_5$ spacetime \eqref{ads5s5}. 
By demanding that the ADM constraint leads to the AdS$_5$-Schwarzschild metric 
when $\alpha\to1$, and denoting by a prime the derivative with respect to $r$, the sum of the momentum and the Hamiltonian constraints in Appendix \ref{app2} reads \cite{Ferreira-Martins:2019svk}
\begin{eqnarray}\label{ctr}
\!{\frac{2G''(r)}{G(r)}}
\!+\!\frac{G^{\prime2}(r)}{G^2(r)}+\frac{2F''(r)}{F(r)}\!-\!
\frac{F^{\prime2}(r)}{F^2(r)}\!+\!\frac{4}{r}\left(\frac{F'(r)}{F(r)}\!-\!\frac{G'(r)}{G(r)}\right)
\!-\!\frac{F'(r)G'(r)}{F(r)G(r)} \!-\!\frac{4G(r)}{r^2}\!=\!k(r,r_{\scalebox{.65}{$0$}},\alpha),
\end{eqnarray}
where the function $k(r,r_{\scalebox{.65}{$0$}},\alpha)$ is given by Eq. (\ref{r1000}) in Appendix \ref{app2}.
The AdS$_5$ boundary resides at $u = 0$, and the horizon is reached in the limit $u \to 1$.
The metric (\ref{1a}) reads
\begin{eqnarray}
\!\!\!\!\!\!\!\!\!d\mathsf{s}^2 = -\frac{r_{\scalebox{.65}{$0$}}^2}{u^2} F(u) {\scalebox{.95}{$\mathrm{d}$}}t^2 + \frac{1}{u^2 G(u)} {\scalebox{.95}{$\mathrm{d}$}}u^2 + \frac{r_{\scalebox{.65}{$0$}}^2}{u^2} \delta_{ij} {\scalebox{.95}{$\mathrm{d}$}}x^i {\scalebox{.95}{$\mathrm{d}$}}x^j. \label{1}
\end{eqnarray}
The standard AdS$_5$-Schwarzschild black brane gets distorted by the ADM embedding, which induces a tidal deformation of the black brane.
Eq. (\ref{ctr}) is compatible with the following blackening factors:
\begin{eqnarray}
F(u) &=& 1 - u^4 + \left (\alpha - 1 \right ) u^6,\label{eq:Nu}\\
G(u) &=& \left (1 - u^4 \right ) \left ( \frac{2 - 3u^4}{2- \left (4\alpha-1\right ) u^4}\right ),
\label{eq:Au}
\end{eqnarray}
where $\alpha\in\mathbb{R}$ is a deformation parameter, which emulates the tidal deformation coefficient of the black brane. In Sec. \ref{rhic} we will investigate how the shear viscosity-to-entropy density ratio can drive a strict specific range of values for $\alpha$, also using the analysis of the experimental data of the QGP at the LHC and the RHIC. 
Here we obtain deformations of the original AdS$_5$-Schwarzschild black brane from embedding on a codimension-one bulk. Ref. \cite{Benini:2006hh} 
obtained deformations of the AdS$_5$-Schwarzschild black brane, implemented by adding flavor branes.

Using the deformed AdS$_5$-Schwarzschild  black brane metric $g_{\mu\nu}$ from the squared line element \eqref{1}, one can take the vector $n_{\nu}=uG(u)\delta_{\nu}^{u}$, which is normal to the boundary, 
with corresponding induced metric $h_{\mu\nu}=g_{\mu\nu}-n_{\mu}n_{\nu}$, endowing the manifold at constant $u$,  given by 
\begin{eqnarray}  \label{inducmetric}
{\scalebox{.95}{$\mathrm{d}$}}s^{2}_{\textsc{ind}}=h_{\mu\nu}{\scalebox{.95}{$\mathrm{d}$}}x^\mu {\scalebox{.95}{$\mathrm{d}$}}x^\nu=-\frac{r_{\scalebox{.65}{$0$}}^{2}}{u^{2}}F(u){\scalebox{.95}{$\mathrm{d}$}}t^{2}+\frac{r_{\scalebox{.65}{$0$}}^{2}}{u^{2}}\delta_{ij}{\scalebox{.95}{$\mathrm{d}$}}x^{i}{\scalebox{.95}{$\mathrm{d}$}}x^{j}.
\end{eqnarray}
Combining the metric \eqref{1}, with coefficients (\ref{eq:Nu}, \ref{eq:Au}), and the GKPW relation \cite{Witten:1998qj, Gubser:1998bc}, the partition function in the dual field theory, can be calculated, from where thermodynamic functions such as the free energy, and entropy, and pressure can be read off. 
Since the partition function can be expressed as
\beq
Z = e^{-S_\textsc{E}},
\eeq
then the Euclidean action 
\begin{align}
\begin{aligned}	\label{Sonshell1}
S_\textsc{E}\!=\!-\frac{1}{16\pi G}&{\int \!{\scalebox{.95}{$\mathrm{d}$}}^{5}x\sqrt{g}\left(R-2\Lambda_5\right)}  -\frac{1}{8\pi G}\;{\lim_{u\to 0}\int \!{\scalebox{.95}{$\mathrm{d}$}}^{4}x\sqrt{h}K}\!+\!S_{\textsc{c.t.}}
\end{aligned}
\end{align}	
has to be evaluated, where the second integral regards the Gibbons--Hawking term (where $K=h^{\mu\nu}\nabla_\mu n_\nu$ 
stands for the trace of the extrinsic curvature), and the last term is a counterterm action to eliminate UV divergences \cite{Emparan:1999pm}. For evaluation of the action \eqref{Sonshell1}, the Euclidean signature is employed, by $t\mapsto i\tau$. Quantum fluctuations of a field in any  curved spacetime induce UV divergences implemented by invariants constructed upon the metric and the curvature in the quantum effective action. 
From the point of view of the dual field theory,  renormalization can be implemented as usual, followed by the addition of a finite number of (local) counterterms to the bare action. Emulating the field-theoretical prescription, the counterterm action in Eq. \eqref{Sonshell1} yields a finite value for the gravitational partition function. This method is also known as the holographic renormalization \cite{deHaro:2000vlm,Natsuume:2014sfa}. 
 Each of the terms in \eqref{Sonshell1} will be computed. For the  Einstein--Hilbert action, the cosmological constant term is written as $-2\Lambda_5=12$ and the scalar curvature can be expanded as  
\begin{eqnarray} \label{scalCurv}
R=-20-8\left(\alpha-1\right)u^{4}+\mathcal{O}(u^8),
\end{eqnarray}
since we are analyzing the region 
$r>r_{\scalebox{.65}{$0$}}$. 
Analogously, the Euclidean metric determinant in Eq. \eqref{Sonshell1} has the form
\beq
	\!\!\sqrt{g}&=&\frac{r_{\scalebox{.65}{$0$}}^4}{u^5}-\frac{\left(\alpha-1\right)}{u}r_{\scalebox{.65}{$0$}}^{4}+\frac{1}{2}(\alpha-1)r_{\scalebox{.65}{$0$}}^{4}u+\frac1{4}{\left(1-\alpha\right)}\left(5+\alpha\right)r_{\scalebox{.65}{$0$}}^{4}u^{3}+\mathcal{O}(u^4).\label{281}
\eeq
Considering the series expansions of both the scalar curvature \eqref{scalCurv} and the metric determinant \eqref{281}, the integral \eqref{Sonshell1} is performed over the product of the expansions, up to ${\scalebox{.93}{$\mathcal{O}$}}(u^8)$. However, only contributions that lead to corrections to order $\alpha^2$ are kept since they constitute the most significant relevance in the numerical analysis. Further corrections at order ${\scalebox{.93}{$\mathcal{O}$}}(\alpha^3)$ imply numerical differences of order $\sim10^{-5}$.
The Einstein--Hilbert term then can be written as 
\begin{align} \label{Ibulk}
\begin{aligned}
{\int \!{\scalebox{.95}{$\mathrm{d}$}}^{5}x\sqrt{g}\left(R-2\Lambda_5\right)}&=\frac12\left(\frac{1}{\epsilon^{4}}-1\right)\left(3{\alpha}-4\right)\left(\alpha-1\right)\!,
\end{aligned}
\end{align}
where the limit $\epsilon\to 0$ is utilized to control divergent terms that shall be canceled out by the counterterm.

On the other hand, the Gibbons--Hawking term $S_\textsc{gh}=
\lim_{u\to 0}\int \!{\scalebox{.95}{$\mathrm{d}$}}^{5}x\sqrt{g}K$ equals a surface term. Therefore we compute it only at the AdS boundary $u\to 0$. 
From the near-boundary expansion of the trace of the extrinsic curvature,  $ 
K=-4\left[1+\left(\alpha-1\right)u^{4}\right]+\mathcal{O}(u^6)$,  
and the induced metric determinant 
\begin{eqnarray} 
\sqrt{h}=r_{\scalebox{.65}{$0$}}^{4}\left[\frac{1}{u^{4}}-\frac{1}{2}+\left(\alpha-1\right)\frac{u^{2}}{2}-\frac{u^{4}}{8}\right]+\mathcal{O}(u^8), \label{hdet}
\end{eqnarray}
the Gibbons--Hawking term reads
\begin{eqnarray}
S_\textsc{gh}=-4\left[\frac{1}{\epsilon^4}-\frac{1}{2}\left(3-2\alpha\right)\right]\ .
\end{eqnarray}
The counterterm in the action \eqref{Sonshell1} cancels power-law and logarithmic divergences and depends solely on the boundary theory. Higher-dimensional terms in the counterterm action that  make a finite contribution to the energy-momentum tensor
are the squares of the Riemann and the Ricci tensors, and the Ricci scalar of the
boundary metric. Therefore the counterterm action reads     \cite{Emparan:1999pm}
\begin{align}
\begin{aligned}\label{ctt}
	\!\!\!\!S_{\textsc{c.t.}}=&\frac{1}{8\pi G}\!\lim_{u\to 0}\int {\scalebox{.95}{$\mathrm{d}$}}^{4}x\sqrt{h}\left\{3+\frac{\mathsf{R}}{4}+\frac{1}{4}\!\left[\mathsf{R}_{\mu\nu}\mathsf{R}^{\mu\nu}\!-\!\frac{4\,\mathsf{R}^{2}}{12}\right]\right\}, 
\end{aligned}
\end{align}
where $\mathsf{R}$ and $\mathsf{R}_{\mu\nu}$ are the scalar curvature and the Ricci tensor constructed upon the induced metric \eqref{inducmetric}, respectively, which vanish for black holes with planar horizon. Therefore only the first term in the counterterm action \eqref{ctt} contributes to thermodynamic quantities, reading 
\begin{eqnarray} 
S_{\textsc{c.t.}}=\frac{3}{8\pi G}\lim_{u\to 0}\int {\scalebox{.95}{$\mathrm{d}$}}^{4}x\sqrt{h}=\frac{3r_{\scalebox{.65}{$0$}}^{4} V \beta}{8\pi G}\left[\frac{1}{\epsilon^{4}}-\frac{1}{2}\right], 
\end{eqnarray}
by Eq. \eqref{hdet}, where $V=\int {\scalebox{.95}{$\mathrm{d}$}}x\,{\scalebox{.95}{$\mathrm{d}$}}y\,{\scalebox{.95}{$\mathrm{d}$}}z$ and $\beta=\int {\scalebox{.95}{$\mathrm{d}$}}\tau$. Summing up the integrals and reinstating the constant factors implies that  the action \eqref{Sonshell1} can be recast as 
\begin{equation} \label{partfnc}
S_\textsc{E}=\frac{V\beta r_{\scalebox{.65}{$0$}}^{4}}{16\pi G}\left({11-15\alpha+3\alpha^{2}}\right).
\end{equation}
The GKPW relation states that Eq.  \eqref{partfnc} represents the partition function associated with  the dual field theory at the AdS boundary. In statistical mechanics, the partition function can be written as $Z=\beta F$, where $F$ denotes the free  energy, yielding thermodynamic functions.

\begin{figure}[H]
\centering\includegraphics[width=7.6cm]{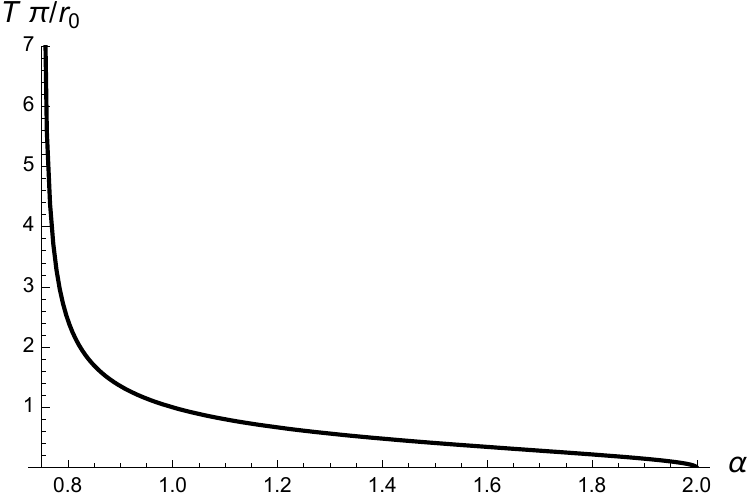}
\caption{\footnotesize Hawking temperature of the deformed AdS$_5$-Schwarzschild black brane as a function of $\alpha$.}
\label{fig:1}
\end{figure}

Temperature is an essential ingredient in computing thermodynamic functions. Since deformed AdS$_5$-Schwarzschild black branes are objects that radiate, in AdS/CFT one can associate them with the Hawking temperature, calculated at the horizon as 	\begin{eqnarray} \label{temperaturegeneral}
	T=\frac{1}{4\pi}\lim_{u\to1}\sqrt{\frac{\dot{g}_{tt}(u)}{\dot{g}_{rr}(u)}}=\frac{r_{\scalebox{.65}{$0$}}}{\pi}\sqrt{\frac{\alpha-2}{3-4\alpha}},
	\end{eqnarray}
for the metric \eqref{1}, where one hereon denotes by the dot the derivative with respect to the coordinate $u$. 
The temperature is related to the parameter of
non-extremality, as extremal black hole solutions, have zero temperature. In the gauge/gravity duality dictionary, a fluid flow in thermal equilibrium at rest has temperature emulated by the  Hawking temperature \eqref{temperaturegeneral}.  
Fig. \ref{fig:1} depicts the temperature of the deformed AdS$_5$-Schwarzschild black brane  (\ref{temperaturegeneral}) with $\alpha$ varying.
\noindent The temperature \eqref{temperaturegeneral} diverges for $\alpha\to 0.75$, and it is a complex number for both the ranges $\alpha > 2$ and $\alpha < 0.75$. For $\alpha = 2$, the temperature of the deformed AdS$_5$-Schwarzschild black brane equals zero and the black brane becomes extremal. Since the deformed AdS$_5$-Schwarzschild  black brane temperature can reach neither imaginary nor divergent values, the deformed AdS$_5$-Schwarzschild  black brane temperature analysis requires the $\alpha$ deformation parameter to reside in the open interval $\alpha\in(0.75,2)$. 
Inverting Eq. (\ref{temperaturegeneral}), one can write the event horizon $r_{\scalebox{.65}{$0$}}$ as a function of the temperature, as 
	\begin{eqnarray} \label{rofT}
	r_{\scalebox{.65}{$0$}}=\pi\sqrt{\frac{3-4\alpha}{\alpha-2}}\,T\ .
	\end{eqnarray} 
Hence the free  energy can be immediately obtained by substituting Eq. \eqref{rofT} in Eq. \eqref{partfnc}, as 
\begin{align}
	\label{freenergyofT}
	\!\!\!\!\!\!\!F=\frac{\pi^{3}V}{16G}\left(11-15\alpha+3\alpha^{2}\right)\left(\frac{3-4\alpha}{\alpha-2}\right)^{2}T^{4}\ .
	\end{align} The thermodynamical state functions can be therefore read off the canonical ensemble. The entropy density, the pressure, and the energy density respectively read
	\begin{eqnarray} \label{entropyofT}
			\!\!\!\!\!\!\!\!\!\!\!\!\!\!s&=&-\frac{1}{V}\frac{\partial F}{\partial T}\!=\!-\frac{\pi^{3}}{4G}\!\left({11\!-\!15\alpha\!+\!3\alpha^{2}}\right)\!\!\left(\frac{3-4\alpha}{\alpha-2}\right)^{\!\!2}\!T^{3},\\
			\!\!\!\!\!\!\!\!P&\!=\!&\!-\frac{\partial F}{\partial V}\!=\!-\frac{\pi^{3}}{16G}\!\left({11\!-\!15\alpha\!+\!3\alpha^{2}}\right)\left(\frac{3\!-\!4\alpha}{\alpha-2}\right)^{2}\!T^{4}\ ,\label{pofT}\\
			\!\!\!\!\!\!\varepsilon &\!=\!& \frac{F}{V}\!-\!Ts \!=\!\frac{5\pi^{3}}{16G}\!\left({11\!-\!15\alpha\!+\!3\alpha^{2}}\right)\!\left(\frac{3-4\alpha}{\alpha-2}\right)^{2}\!T^{4}\label{espsofT}.	\end{eqnarray}
			It is worth emphasizing that the pressure exhibits the relativistic Stefan--Boltzmann pattern, governed by temperature to the fourth power. 	The energy density of the dual fluid, in the hydroelastic complementarity in black branes, corresponds to 
			the height function encoding the deformation of an elastic membrane \cite{Emparan:2016sjk}. 	
			Besides, the specific heat at constant volume can be written as \beq
			\!\!\!\!\!\!C_{\scalebox{.55}{$V$}}&=&T\left(\frac{\partial s}{\partial T}\right)_{\scalebox{.55}{$V$}}=-\frac{3\pi^{3}}{4G}\!\left({11\!-\!15\alpha\!+\!3\alpha^{2}}\right)\!\!\left(\frac{3-4\alpha}{\alpha-2}\right)^{\!\!2}\!T^{3}.
\eeq
The dual field theory on the boundary inherits the thermodynamics from the deformed AdS$_5$-Schwarzschild black brane, on the gravity side. 
The energy-momentum tensor
\begin{equation}
T^{\mu\nu}=\left(\varepsilon+P\right)u^{\mu}u^{\nu}+Pg^{\mu\nu},\label{emt}
\end{equation} in the dual QFT, is encoded in the dual gravitational sector. The gauge invariant coupling comprised by the trace of Eq. (\ref{emt}) can be calculated 
 using Eqs. (\ref{pofT}, \ref{espsofT}) evaluated at the AdS boundary,  reading
\beq
\!\!\!\!\!\!\!\!\! g_{\mu\nu}T^{\mu\nu}&\!=\!&-\varepsilon+3P=-\frac{\pi^{3}}{2G}\left({11-15\alpha+3\alpha^{2}}\right)\!\left(\frac{3-4\alpha}{\alpha-2}\right)^{\!\!2}T^{4}.\label{trEMT}
\eeq
For computing $\eta/s$ it is more useful to employ Eq. (\ref{rofT}) and write Eq. \eqref{entropyofT} in terms of the horizon, as 
\bltx{\begin{eqnarray} \label{sofr}
s=-\frac{r_{\scalebox{.65}{$0$}}^{3}}{4G}\left(11-15\alpha+3\alpha^{2}\right)\left(\frac{3-4\alpha}{\alpha-2}\right)^{1/2} \,,
\end{eqnarray}}
and to express the surface area of the deformed AdS$_5$-Schwarzschild black brane as $A=4GsV$.



\section{Shear viscosity-to-entropy density ratios of deformed AdS$_5$-Schwarzschild black branes}
\label{scru}\label{sec:eta_s_ads_mgd_5}

AdS/CFT is a paradigm relating gravity in AdS spacetime to a large-$N_c$ CFT, located on the AdS boundary. Perturbatively, considering a $1/N_c$ expansion, quantum fields in the AdS bulk correspond to CFT operators \cite{Rangamani:2009xk,Bhattacharyya:2008jc}. The dynamics of Einstein's effective equations, describing weakly-coupled gravity in an AdS space, rules the  dynamics of the energy-momentum tensor of strongly coupled QFT on the AdS boundary 
\cite{Witten:1998qj,Gubser:1998bc}. Therefore this setup can be used to investigate deformed AdS$_5$-Schwarzschild  black branes \eqref{1}, with coefficients (\ref{eq:Nu}, \ref{eq:Au}), and derive the ${\eta}/{s}$ ratio in this gravitational background. 
Consider an off-diagonal bulk perturbation $h_{xy}$ 
\begin{equation}\label{ref12}
 {\scalebox{.95}{$\mathrm{d}$}}s^2 = {\scalebox{.95}{$\mathrm{d}$}}\mathsf{s}^2 + 2h_{xy} {\scalebox{.95}{$\mathrm{d}$}}x {\scalebox{.95}{$\mathrm{d}$}}y \ ,
\end{equation}
\noindent where $d\mathsf{s}^2$ is the deformed AdS$_5$-Schwarzschild black brane metric in Eq. \eqref{1}, with coefficients (\ref{eq:Nu}, \ref{eq:Au}). Ref. \cite{Ferreira-Martins:2019svk} demonstrated that this perturbation propagates at the speed of light, corroborating a stable causal structure. 
 One can remember Eq. \eqref{oooo}, where $h_{xy}^{\scalebox{.65}{$(0)$}}$ regards the perturbation to the dual field theory on the boundary. The term $h_{xy}^{\scalebox{.65}{$(0)$}}$ is the solution of the Laplace-Beltrami operator equation for the massless scalar field perturbation ${\scalebox{.98}{$\upvarphi$}}\equiv g^{xx}h_{xy}$  \cite{kss}
\begin{equation}
	\frac{1} {\sqrt{-g}}\nabla_{\scalebox{.6}{$A$}} \left( \sqrt{-g} g^{\scalebox{.6}{$AB$}} \nabla_{\scalebox{.6}{$B$}} {\scalebox{.98}{$\upvarphi$}} \right) = 0,
\end{equation} which is asymptotically related to $h_{xy}$  by 
the expression
\begin{equation}
 g^{xx}h_{xy} \sim h_{xy}^{\scalebox{.65}{$(0)$}}\, \left(1 \!+\! h_{xy}^{\scalebox{.65}{$(1)$}} u^4 \right).
 \label{eq:perturb_asym_def}
\end{equation}
\noindent 
The family of deformed AdS$_5$-Schwarzschild  black branes asymptotically behaves analogously to the standard AdS$_5$-Schwarzschild black brane \eqref{eq:sads_5_u}, whose asymptotic limit is given by the metric \eqref{eq:asym_ads}. The term ${\scalebox{.98}{$\upvarphi$}}=g^{xx}h_{xy}$ emulates an external source of the $\tau^{xy}$ boundary operator. The metric is dual to the quantum field theory energy-momentum tensor. Hence, the shear viscosity
can be computed by adding metric fluctuations on the gravity side of the correspondence to the deformed AdS$_5$-Schwarzschild  black brane. From the effective action for ${\scalebox{.98}{$\upvarphi$}}$, one can then derive the retarded two-point function
for the $\tau^{xy}$ component of the boundary CFT energy-momentum tensor, and read off $\eta$ in the small ${\bm{\mathfrak{q}}}$ and $\omega$
limits, as in Eq. (\ref{xxx}). 
One can calculate the response $\updelta \left \langle \tau^{xy} \right \rangle$ using Eq. \eqref{eq:responsegeneralgkpw} as 
\begin{equation}
\updelta \left \langle \tau^{xy} \right \rangle = \frac{r_{\scalebox{.65}{$0$}}^4}{16 \pi G}4 h_{xy}^{\scalebox{.65}{$(1)$}}h_{xy}^{\scalebox{.65}{$(0)$}} \ , 
\label{eq:response_2}
\end{equation}
\noindent where the $1/16 \pi G$ factor was reintroduced. When Eq. \eqref{oooo} is compared to Eq.  \eqref{eq:response_2} it wits  
\begin{equation} \label{eq:eta_quase}
 i \omega \eta = \frac{r_{\scalebox{.65}{$0$}}^4}{4\pi G} h_{xy}^{\scalebox{.65}{$(1)$}} .
\end{equation}
Therefore taking the shear viscosity in Eq. \eqref{eq:eta_quase} and the entropy density \eqref{sofr}, one obtains the shear viscosity-to-entropy density ratio for the deformed AdS$_5$-Schwarzschild black brane:
\begin{eqnarray}
\!\!\!\!\!\!\!\!\!\!\!\!\!\!\frac{\eta}{s}=-\frac{r_{\scalebox{.65}{$0$}}}{(11-15\alpha+3\alpha^{2})\pi}\left(\frac{\alpha-2}{3-4\alpha}\right)^{1/2}\frac{h_{xy}^{\left(1\right)}}{i\omega},
\label{eq:eta_s_geral}
\end{eqnarray}
\noindent When stationary perturbations ${\scalebox{.98}{$\upvarphi$}}(u,t) = \upphi(u) e^{-i\omega t}$ are regarded, the following ODE governs $\upphi(u)$:
\begin{equation}
\ddot\upphi(u)+ \frac12\left ( \frac{\dot{F}(u)G(u)}{2 F(u)}+\frac{F(u)\dot{G}(u)}{G(u)} - \frac{3}{u}\right ) \dot\upphi(u) + \frac{\omega^2}{F(u)G(u)r_{\scalebox{.65}{$0$}}^2} \upphi(u) = 0.
\label{eq:pertu_edo2}
\end{equation}
The solution of Eq. \eqref{eq:pertu_edo2}
can be obtained by imposing two boundary conditions. The first one comprises the incoming wave in the near-horizon ($u \rightarrow 1$) region, whereas the second one is the boundary condition of the first type at the AdS boundary, $\lim_{u\rightarrow 0}\upphi = \upphi^{\scalebox{.65}{$(0)$}}$, for  
\beq \label{hxy0}
h_{xy}^{\scalebox{.65}{$(0)$}} =  \upphi^{\scalebox{.65}{$(0)$}} e^{-i \omega t}.\eeq 
\noindent The near-horizon incoming-wave boundary condition can be derived when Eq. \eqref{eq:pertu_edo2} is solved for  $u \rightarrow 1$, corresponding to the solution 
\begin{eqnarray}\label{eq:sol_nh}
 \upphi(u) \propto \exp \left (\pm i \frac{\omega}{r_{\scalebox{.65}{$0$}}} \sqrt{\frac{4\alpha -3}{\alpha - 1}}\sqrt{1-u}\right).
\end{eqnarray}
In tortoise coordinates, Eq. \eqref{eq:sol_nh} is a plane wave solution \cite{Natsuume:2014sfa}. The positive [negative] sign in the exponent corresponds to an outgoing [ingoing] wave. The near-horizon $u\to1$ boundary condition then permits one to fix the 
negative sign in Eq. \eqref{eq:sol_nh}.

Eq. \eqref{eq:pertu_edo2} is now solved in the interval  $0\leq u\leq 1$ as a power series in $\omega$. As the hydrodynamic limit of this solution is taken into account, it is enough to regard linear order, 
\begin{equation}
 \upphi(u) = \Phi_0(u) + \omega \Phi_1(u) +\mathcal{O}(\omega^2)\ .
 \label{eq:sol_omega_power}
\end{equation}
The second term in Eq. \eqref{eq:pertu_edo2} can be neglected due to its order $\mathcal{O}(\omega^2)$. Integration yields
\begin{equation}
	\Phi_a(u) = c_a + k_a \int^u {\scalebox{.95}{$\mathrm{d}$}} \textsf{u}\,\frac{\textsf{u}^3}{\sqrt{F(\textsf{u})G(\textsf{u})}}\ , 
\end{equation}
where $c_a$ and $k_a$ are integration constants, and $a=0,1$. Eq. \eqref{eq:sol_omega_power} therefore implies that  
\begin{equation}
 \upphi(u) = c_0 + \omega c_1 + \left (k_0 + \omega k_1 \right ) \int^u {\scalebox{.95}{$\mathrm{d}$}} \textsf{u}\,\frac{\textsf{u}^3}{\sqrt{F(\textsf{u})G(\textsf{u})}} \ .
 \label{eq:sol_omega_power_general}
\end{equation}
To impose boundary conditions we solve the integral (\ref{eq:sol_omega_power_general}) in the asymptotic region $u\rightarrow 0$ and also near-horizon, $u\rightarrow 1$, respectively yielding, up to leading-order terms, the expressions
\begin{eqnarray}
\lim_{u \rightarrow 0}\upphi(u) &=& c_0 + \omega c_1 + \left (k_0 + \omega k_1 \right )\frac{u^4}{4},\\
\lim_{u \rightarrow 1}\upphi(u) &=& c_0 + \omega c_1 + \left (k_0 + \omega k_1 \right )\frac{3-4\alpha}{\alpha-1}\sqrt{\frac{\alpha-1}{3-4\alpha}}\sqrt{1-u}.
\end{eqnarray}
The Dirichlet boundary condition fixes the first pair of integration constants
\begin{eqnarray}
\!\!\!\!\!\!\!\!\!\!\!\!\!\! \lim_{u\rightarrow0} \upphi(u) =c_0 + \omega c_1 + \left (k_0 + \omega k_1 \right ) \lim_{u\rightarrow0} \frac{u^4}{4} =  \upphi^{\scalebox{.65}{$(0)$}},
\end{eqnarray} 
implying that $c_0 + \omega c_1 = \upphi^{\scalebox{.65}{$(0)$}}$. The near-horizon limit wits 
\begin{equation}
   \upphi(u) \approx \upphi^{\scalebox{.65}{$(0)$}} - \left (k_0 + \omega k_1 \right ) \frac{(4\alpha-3)}{\alpha-1}\sqrt{\frac{\alpha-1}{4\alpha-3}}\sqrt{1-u}.
   \label{eq:sol_general_nh}
\end{equation}
 Eq. \eqref{eq:sol_nh} can be expanded up to ${\scalebox{.93}{$\mathcal{O}$}}(\omega^2)$ and using Eq. \eqref{eq:sol_general_nh}  to fix the proportionality constant yields
\begin{equation}
   \upphi(u) \approx \upphi^{\scalebox{.65}{$(0)$}}\left(1 - i \frac{\omega}{r_{\scalebox{.65}{$0$}}}\sqrt{\frac{4\alpha-3}{\alpha-1}}\sqrt{1-u}\right).
   \label{eq:sol_nh_power_omega}
\end{equation}
Comparing Eq. \eqref{eq:sol_general_nh} to  \eqref{eq:sol_nh_power_omega} fixes the second pair of integration constants and 
yields the solution to read 
\begin{equation}
 \!\!\!\upphi(u) \!=\! \upphi^{\scalebox{.65}{$(0)$}} \left ( 1 \!+\! i \frac{\omega}{r_{\scalebox{.65}{$0$}}} \frac{{\rm sign}(\alpha -1)}{{\rm sign}(4\alpha\!-\!3)} \int^u\, {\scalebox{.95}{$\mathrm{d}$}} \mathsf{u}\frac{\mathsf{u}^3}{\sqrt{F(\mathsf{u})G(\mathsf{u})}} \right ) \ . 
\end{equation}
Hence, the time-dependent perturbation is asymptotically given by
\begin{equation}
	{\scalebox{.98}{$\upvarphi$}}(u) =  
	\upphi(u) e^{-i\omega t} \sim  e^{-i\omega t}\upphi^{\scalebox{.65}{$(0)$}} \left ( 1 + i \frac{\omega}{r_{\scalebox{.65}{$0$}}} \frac{{\rm sign}(\alpha -1)}{{\rm sign}(4\alpha\!-\!3)} \frac{u^4}{4}\right ) \ .
\label{eq:perturb_asym_solution}
\end{equation}
Eq. (\ref{eq:perturb_asym_def}) together with Eq. (\ref{eq:perturb_asym_solution}) then yield 
\begin{equation}
h_{xy}^{\scalebox{.65}{$(1)$}} = \frac{i\omega}{4r_{\scalebox{.65}{$0$}}} \frac{{\rm sign}(\alpha -1)}{{\rm sign}(4\alpha\!-\!3)}\ ,
\label{eq:key}
\end{equation}
where it is here useful to remember Eq. (\ref{hxy0}). 
The value $\alpha = 1$ is relevant here, since in this case,  the standard AdS$_5$-Schwarzschild black brane \eqref{eq:sads_5_u} is recovered. 

\begin{figure}[H]
\centering\includegraphics[width=8.5cm]{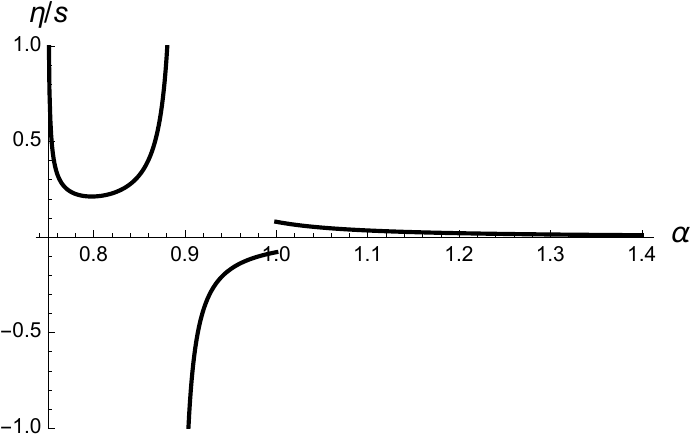}
\caption{\footnotesize The $\frac{\eta}{s}$ ratio of the deformed AdS$_5$-Schwarzschild  black brane as a function of $\alpha$.}
\label{fi2}
\end{figure}

Eq. \eqref{eq:key} can now be substituted into Eq. \eqref{eq:eta_s_geral}, yielding 
\begin{equation} \label{eq:etaSfinal}
\frac{\eta}{s}=\frac{1}{4\pi}\left(\frac{{\rm sign}(\alpha -1)}{11-15\alpha+3\alpha^{2}}\right)\left(\frac{\alpha-2}{3-4\alpha}\right)^{1/2}.\end{equation}	
As expected, the $\eta/s$ ratio is independent of the black brane event horizon. 
Fig. \ref{fi2} portrays $\eta/s$ for the deformed AdS$_5$-Schwarzschild black brane in Eq. (\ref{eq:etaSfinal}) as a function of $\alpha$.
\noindent In the limit $\alpha \to1$, the deformed AdS$_5$-Schwarzschild  black brane ${\eta}/{s}$ ratio equals ${1}/{4\pi}\approxeq7.95774\times 10^{-2}$, which recovers the KSS viscosity-bound saturation of the standard AdS$_5$-Schwarzschild   black brane \cite{Buchel:2004di}. It is worth mentioning that when both the large-$N_c$ limit and the large ’t Hooft coupling $\lambda=g_{\scalebox{.65}{\textsc{YM}}}^2N_c\to\infty$ are  not demanded, the shear viscosity-to-entropy density ratio of the standard AdS$_5$-Schwarzschild   black brane, corresponding to the limit $\alpha\to1$, can be written as \cite{Buchel:2004di,Buchel:2008ae}
\begin{equation}
\label{nccc}
\frac{\eta}{s}
=
\frac{1}{4\, \pi }
\left[1 + \frac {15\,\zeta(3)}{\lambda^{3/2}} +\frac{5\sqrt{\lambda}}{16N_c^2} +\mathcal{O}\left[\exp\left(-\frac{8\pi^2N_c}{\lambda}\right)\right] \right]
\ ,
\end{equation} for $\zeta (3)$ denoting the Ap\'{e}ry constant, and the next-to-leading order term regards 
the first stringy correction to GR.  In the regime of large of values of $N_c$ and large 't Hooft coupling
$g_{\scalebox{.65}{\textsc{YM}}}^2N_c\to\infty$,  one has the KSS viscosity-bound saturation ${\eta}/{s}={1}/{4\pi}$. 
Deforming the AdS$_5$-Schwarzschild black brane can open new spots to compare the thermodynamics of strongly-coupled QCD-like gauge theories with lattice and experimental results of the QGP. Specific features of the bulk do not need to be addressed, as the Campbell--Magaard embedding theorem ensures the existence of Einstein's field equations solutions 
 \cite{Ferreira-Martins:2019svk}.

The KSS viscosity-bound saturation is a compelling apparatus to scrutinize strongly-interacting
quantum field theoretical systems, like the QGP. Fig. \ref{fi2} shows the divergence of ${\eta}/{s}$ for $\alpha \to 0.9^+$.
Eq. \eqref{eq:etaSfinal} makes it clear that $\displaystyle{\lim_{\alpha \to 2}\eta/s=0}$. For $\alpha \to 1.4$, $\eta/s \approx 0.0089$ corresponds to $\sim 11\%$ the standard KSS viscosity-bound saturation for the AdS$_5$-Schwarzschild black brane.
Fig. \ref{fi2} illustrates that in the allowed range $0.75<\alpha < 0.9$ the KSS viscosity-bound saturation is not violated. The range  $0.9\leq\alpha < 1$ implies ${\eta}/{s}<0$, which indicates some instability against perturbations triggering contraction or expansion. The KSS viscosity-bound saturation ${\eta}/{s} = {1}/{4\pi}$ implies $\alpha = 1$, recovering the standard AdS$_5$-Schwarzschild black brane (\ref{eq:sads_5_u}). A first ancillary consistency test can be implemented, taking into account the metric component  (\ref{eq:Nu}), which defines the deformed AdS$_5$-Schwarzschild black brane event horizon. Let us denote by $u_\alpha = r_{\scalebox{.65}{$0$}}/r_\alpha$ the solution of  the algebraic equation $F(u)=0$, in Eq. (\ref{eq:Nu}). The parameter $\alpha$ must also be chosen in a physically consistent way, to comply with a real event horizon, requiring the equation $F(u)=0$ to have at least one real solution. For it to occur, the formal range $1 < \alpha \leq 2$ must be restricted to $1 < \alpha \leq 1.3841$. 
A second auxiliary consistency test imposes that the event horizon  $r_{\scalebox{.65}{$0$}}=\lim_{\alpha\to1} r_\alpha$ is a Killing horizon, further restricting the range to $1 < \alpha \lesssim  1.21$, complying with a causal spacetime and consistent thermodynamic state functions. Hence, the deformed AdS$_5$-Schwarzschild black brane $\alpha$ parameter lies in the ranges
\begin{equation}\label{ra22}
0.75< \alpha < 0.90\;\;\;\;\;\;\;\text{and}\quad 1 < \alpha \lesssim 1.21.
\end{equation}
In the last range, the KSS viscosity bound is violated, suggesting that the deformed AdS$_5$-Schwarzschild black brane solution can be complementarily obtained by 
a gravitational action containing higher-order curvature terms. This is an alternative manner for obtaining the metric of the deformed AdS$_5$-Schwarzschild black brane  \eqref{1}, with coefficients (\ref{eq:Nu}, \ref{eq:Au}), which was derived through the embedding protocol. We address this possibility in the Appendix \ref{app3}. It is not the first example in the literature of a setup that violates the KSS viscosity bound and does not involve higher-order derivative theories of gravity, in the gauge/gravity correspondence. The strongly-coupled $\mathcal{N} = 4$ super-Yang--Mills plasmas can describe pre-equilibrium stages of the QGP, whose $\eta/s$, with shear viscosity transverse to the direction of anisotropy, was shown to saturate the KSS viscosity bound in Ref. \cite{Rebhan:2011vd}. Besides, anisotropy in the shear viscosity induced by external magnetic fields in strongly-coupled plasmas makes $\eta/s$  violate the KSS viscosity bound \cite{Critelli:2014kra}. 
Also, spatially anisotropic $\mathcal{N} = 4$ super-Yang--Mills theory at finite chemical potential was approached in  Ref. \cite{Ge:2014aza}, with longitudinal shear viscosity for prolate anisotropy violating the KSS viscosity bound. 
Anisotropic black branes with a dilaton were addressed, with the KSS saturation bound violated \cite{Jain:2014vka}. 
Other more general scenarios where the KSS viscosity bound is violated were reported in Refs. \cite{Sadeghi:2022bsh,Sadeghi:2019muh,Sadeghi:2018ylh}. 
Ref.  \cite{Romatschke:2007mq} used Glauber-type initial conditions with the minimum bias of the integrated elliptic flow coefficient, to study the QGP, also leading to KSS viscosity-bound violation, although it might be circumvented with the color-glass condensate initial conditions \cite{Karapetyan:2017edu}. 

\begin{figure}[H]
\centering\includegraphics[width=8cm]{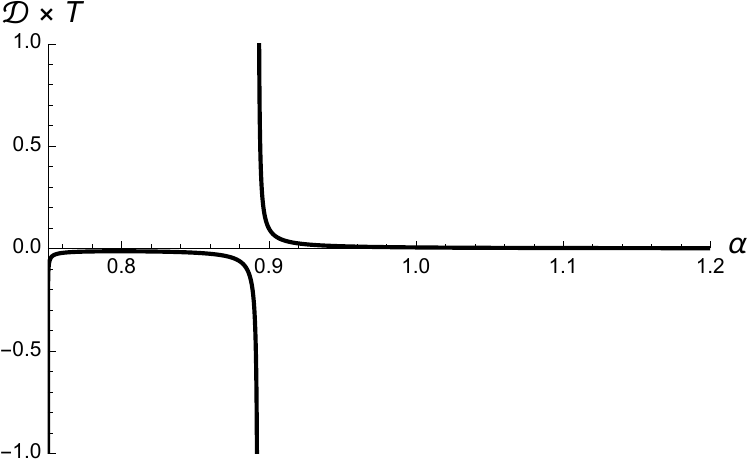}
\caption{\footnotesize Shear mode damping constant times the temperature 
 of the deformed AdS$_5$-Schwarzschild  black brane as a function of $\alpha$.}
\label{fi233}
\end{figure}

Deformed AdS$_5$-Schwarzschild  black branes can have an analog hydrodynamic description. The approach to their gravitational perturbations indicates that fluctuations of the stretched horizon have to encode the complete collection of hydrodynamical modes. Methods in fluid/gravity correspondence must replicate wave modes describing the sound propagation and non-propagating modes including shear and diffusion \cite{Dudal:2018rki}.
Hydrodynamics, as an effective theory, designates the dynamics ruling any thermal system whose time and length scales are large, contrasted to microscopic scales. 
The dual hydrodynamical description of black branes makes the shear diffusion mode emerge, whose diffusion constant is proportional to the shear viscosity  \cite{Iqbal:2008by}. 
The shear mode dispersion relation reads  $\omega=\omega( \mathfrak{q})= -i\mathcal{D}  \mathfrak{q}^2$, which arises from the lowest pole in the long-wavelength, low-frequency hydrodynamic limit, due to the graviton absorption upon gravitational perturbations. The universality of $\eta/s$ strongly depends on the shear mode transforming as
a helicity two state under the rotational symmetry \cite{Erdmenger:2010xm}. 
The diffusion can then be interpreted as a byproduct of the viscosities of the
dual gauge theory plasmas also in the limit $ \mathfrak{q}\to0$, where the shear mode damping constant can be expressed by  
$\mathcal{D}=\eta/(\upepsilon+P)$ \cite{Herzog:2002fn}. 
 Using the thermodynamic functions (\ref{entropyofT}) -- (\ref{espsofT}) and Eq. \eqref{eq:etaSfinal}, one 
 finds  the transverse diffusivity given by the shear mode damping constant as
 \beq\label{eq:etaSfinal1}
 \mathcal{D} = \frac{\eta}{\epsilon + P} = -\frac{1}{2 \pi ^4 \left(11-15 \alpha +3
   \alpha ^2\right)T}\sqrt{\frac{\alpha -2}{3-4 \alpha }} .
 \eeq 
\textcolor{black}{Fig. \ref{fi233} illustrates Eq. (\ref{eq:etaSfinal1}) as a function of $\alpha$.}
 \noindent When ${\alpha\to1}$, the shear mode damping diffusivity constant  $\mathcal{D} = 2.9485\times 10^{-15}\; {\rm m^2/s}$ can be recovered, 
for the confinement-deconfinement phase transition temperature of the QGP.

It is worth mentioning that Ref. \cite{daRocha:2021xwq} approached phenomenological aspects of deformed AdS$_5$-Schwarzschild  black branes in AdS/QCD, using the holographic entanglement entropy as a representative of the order parameter that governs the confinement-deconfinement phase transition.  
A hadronic confined phase can undergo a phase transition, at the Hagedorn temperature, to the deconfined one. 
The deconfinement temperature range $122\,{\rm MeV}\lesssim T_{\textsc{h}}\lesssim 170$ MeV was obtained in the AdS/QCD hard-wall model and lattice QCD as well \cite{Cucchieri:2007rg}, whereas  the range  $190\,{\rm MeV}\lesssim T_{\textsc{h}}\lesssim 200$ MeV was obtained in the AdS/QCD soft-wall model. The value $T_{\textsc{h}} \approx 175\pm 15$ MeV was obtained from the unquenched glueball spectrum fit in Ref. \cite{Afonin:2018era} and some other relevant results of the confinement-deconfinement
temperature were addressed in Refs. \cite{Dudal:2017max,Braga:2020opg}. 
 The HotQCD Collaboration found $T_{\textsc{h}} = 156.5 \pm 1.5$ MeV \cite{HotQCD:2018pds} and the most recent result for the transition temperature is $T_{\textsc{h}} = 158.0\pm0.6$ MeV \cite{Borsanyi:2020fev}. The range $
0.878\lesssim \alpha\lesssim 1.192$ 
implementing the black brane metric \eqref{1}, with coefficients (\ref{eq:Nu}, \ref{eq:Au}), as a mild deformation of the standard AdS$_5$-Schwarzschild black brane, matches the predictions by both the AdS/QCD hard wall and soft wall models, including the most recent ones by lattice QCD  \cite{daRocha:2021xwq}. Therefore, one can summarize the intersection of this last range for $\alpha$ with the ranges in \eqref{ra22} to conclude that, up to now, we found the reliable ranges 
\beq\label{ra33}
0.878\lesssim\alpha\lesssim 0.900,\qquad\qquad 1\leq \alpha \lesssim 1.192,
\eeq
allowed for deformations of the standard AdS$_5$-Schwarzschild black brane. In the next section, the 
analysis of the experimental data regarding the temperature-dependent $\eta/s$ of the QGP \cite{JETSCAPE:2020shq,Bernhard:2019bmu,Parkkila:2021yha,Nijs:2022rme,Nijs:2022rme}
 will be used to bind more strictly the allowed ranges of the deformation parameter $\alpha$, also incorporating the dependence of the temperature of the QGP. In particular, we will show that the range $0.878\lesssim\alpha\lesssim 0.900$ is experimentally forbidden.

\section{Constraining black brane deformations from experimental data of the QGP and heavy-ion collisions 
at LHC and RHIC}
\label{rhic}

One of the most relevant results from heavy-ion collisions at RHIC is the experimental production of QCD hot matter in  ultrarelativistic heavy-ion collisions,  behaving nearly as an ideal fluid flow, instead of a gas constituted by quarks and gluons. 
The QGP shear viscosity has measurable effects on the momentum distribution of hadrons produced in heavy-ion collisions. 
The shear viscosity reduces asymmetries in the hadron momentum distributions, in heavy-ion collisions, generated by anisotropic gradients of pressure in the QGP, which are induced by spatial inhomogeneities emerging from quantum fluctuations at the nuclear impact \cite{Rougemont:2015ona}. 
The QGP can be categorized by macroscopic properties, including transport and response coefficients, obtained in heavy-ion collisions at LHC and RHIC for temperatures  $150\;{\rm MeV} \lesssim T\lesssim 350$ MeV. 
The QGP macroscopic features reflect the underlying hydrodynamical fluid microscopic interactions governed by QCD. The strongly-coupled microscopic dynamics of QGP prevent any perturbative approach \cite{JETSCAPE:2020shq}. 
Hadronic measurements obtained at the LHC and RHIC severely constrain the temperature dependence of the shear viscosity-to-entropy density ratio of the QGP. 
Hydrodynamical simulations data analysis regarding the QGP produced from heavy-ion collisions at LHC and RHIC indicate a very small value of the shear viscosity-to-entropy ratio for the QCD plasma \cite{Kharzeev:2007wb}. The origin of this minimal value can be sourced by the strongly-coupled $\mathcal{N} = 4$ super-Yang--Mills theory in holographic AdS/CFT, which points out that the ratio $\eta/s$ of the QGP, at the strong-coupling regime, attains small values.

\begin{figure}[H]\begin{center}
\includegraphics[scale=0.65]{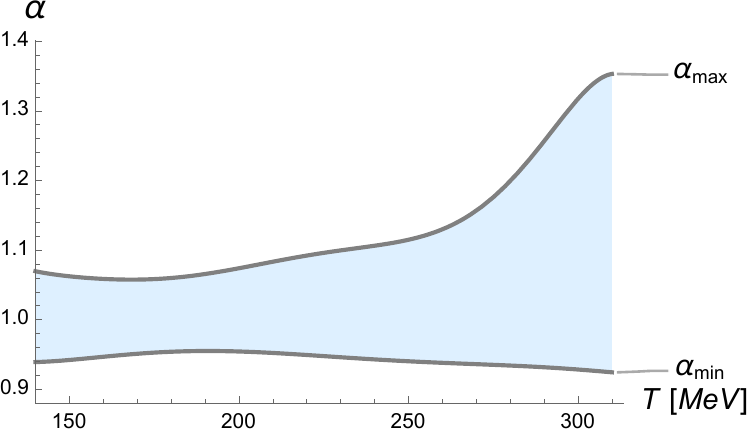}
\caption{\footnotesize Range of the parameter $\alpha$ in the deformed AdS$_5$-Schwarzschild black brane \eqref{1}, with coefficients (\ref{eq:Nu}, \ref{eq:Au}), as a function of the QGP temperature (MeV), using the analysis of experimental data of the QGP by the JETSCAPE Bayesian model \cite{JETSCAPE:2020shq} for $\eta/s$.}
\label{gmm22}\end{center}
\end{figure}

We can consider the analyses of the latest experimental estimates for the QGP shear viscosity-to-entropy density ratio \cite{JETSCAPE:2020shq,Parkkila:2021yha,Bernhard:2019bmu,Nijs:2022rme,Nijs:2020ors,Yang:2022ixy}. To accomplish this task, the first analysis regards the JETSCAPE Bayesian model   \cite{JETSCAPE:2020shq}. We can employ the shear viscosity-to-entropy density ratio \eqref{eq:etaSfinal} obtained to the deformed AdS$_5$-Schwarzschild  black brane metric solution \eqref{1}, with coefficients (\ref{eq:Nu}, \ref{eq:Au}),
and compare it to the experimental values of these ratios measured for the QGP measured at LHC and RHIC. 
Fig. \ref{gmm22} illustrates the bound on the 
parameter $\alpha$ in the deformed AdS$_5$-Schwarzschild  black brane metric solution as a function of the QGP temperature (MeV), using the results by the  JETSCAPE Bayesian model. The range of $\alpha$ is severely restricted when 
the shear viscosity-to-entropy density \eqref{eq:etaSfinal} of the deformed AdS$_5$-Schwarzschild  black brane is compared to the data in the right panel of Fig. 1 in Ref. \cite{JETSCAPE:2020shq}, where Bayesian inference 
converted theoretical and experimental uncertainties into probabilistic bounds for the shear  viscosity. 
\noindent 
Regarding the Hagedorn temperature, the range $0.941\lesssim \alpha\lesssim 1.072$ for the parameter regulating deformations of the AdS$_5$-Schwarzschild black brane is reliable. 
The extremal values of the bound for $\alpha$ in Fig. \ref{gmm22} can be interpolated, with the explicit expressions:
\begin{subequations}
\beq
\alpha_{\textsc{min}}(T)&=&1.12258\times 10^{-16} T^7-1.28334\times 10^{-13} T^6+4.70550\times 10^{-11}
   T^5+8.56005\times 10^{-10} T^4\nonumber\\
   &&\qquad\,-3.80774 \times 10^{-6} 
   T^3+1.04433\times 10^{-3}  T^2-1.13888 T+5.50888,\\
     \alpha_{\textsc{max}}(T)&=&-5.34789\times 10^{-15} T^7+7.77042\times 10^{-12} T^6-4.76218\times 10^{-9}
   T^5+1.59678\times 10^{-6} T^4\nonumber\\
   &&\qquad\,+3.16640\times 10^{-4}
   T^3+3.71828\times 10^{-2} T^2-2.39851\,T+66.6968,
   \eeq\end{subequations}
within $10^{-3}\%$ interpolation error. 

\begin{figure}[H]\begin{center}
\includegraphics[scale=0.65]{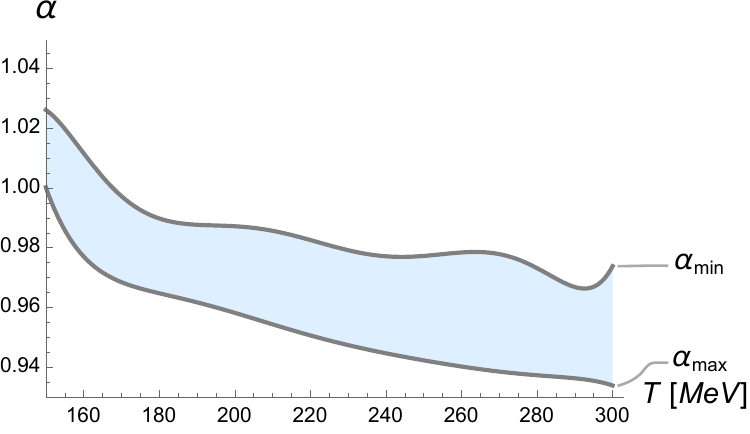}
\caption{\footnotesize \footnotesize Range of the parameter $\alpha$ in the deformed AdS$_5$-Schwarzschild black brane \eqref{1}, with coefficients (\ref{eq:Nu}, \ref{eq:Au}),
 as a function of the QGP temperature (MeV),  using the analysis of experimental data of the QGP at LHC and RHIC by the Duke group \cite{Bernhard:2019bmu} for $\eta/s$.}
\label{gmm45}\end{center}
\end{figure}

 The second similar result can be deployed by considering the experimental data for the shear viscosity-to-entropy density ratio of the QGP, obtained by the Duke group \cite{Bernhard:2019bmu}. In this analysis, $\eta/s$ was estimated from heavy nuclei collision data at $\sqrt{s_{NN}} =  2.76$ and 5.02 TeV,  producing transient
droplets of QGP, whose spacetime evolution was scrutinized by Bayesian parameter estimation methods. 
Effects of the pre-hydrodynamic evolution on final-state observables, in heavy-ion
collisions were addressed in Ref. \cite{NunesdaSilva:2020bfs}. 
Taking into account these results, 
 the shear viscosity-to-entropy density ratio \eqref{eq:etaSfinal} for the deformed AdS$_5$-Schwarzschild  black brane metric solution \eqref{1}, with coefficients (\ref{eq:Nu}, \ref{eq:Au}), can be equaled to the experimental values of $\eta/s$ of the QGP and algebraically solved for the parameter $\alpha$. In particular, the estimated temperature-dependent value of $\eta/s$ of the QGP, in the top panel of Fig. 2 of Ref. \cite{Bernhard:2019bmu}, is compared to the shear  viscosity-to-entropy density ratio in Eq. \eqref{eq:etaSfinal}, and solved to obtain the minimum and maximum values of $\alpha$. The temperature-dependent bound on $\alpha$ is plotted in Fig. \ref{gmm45}.
\noindent 
At the Hagedorn temperature, one can extract the range $0.980\lesssim \alpha\lesssim 1.014$ for the parameter regulating deformations of the AdS$_5$-Schwarzschild black brane. 
We can interpolate the extremal values of the bound for $\alpha$ in Fig. \ref{gmm45} by the following polynomials:
\begin{subequations}
\beq\label{dukee1}
\alpha_{\textsc{min}}(T)&=&1.31552\times 10^{-14} T^7-2.05027\times 10^{-11} T^6+1.35627\times 10^{-8}
   T^5-4.9345\times 10^{-6} T^4\nonumber\\
   &&\qquad\,+1.06604\times 10^{-3}
   T^3-0.13669\times 10^{-1} T^2+9.62617T-286.074,\\
  \alpha_{\textsc{max}}(T)&=&-2.91826\times 10^{-15} T^7+4.80814\times 10^{-12} T^6-3.37657\times 10^{-9}
   T^5+1.30981\times 10^{-6} T^4\nonumber\\
   &&\qquad\,+3.03006 \times 10^{-4} 
   T^3+4.17916\times 10^{-2}  T^2-3.18149 T-104.100,\label{dukee2}
   \eeq\end{subequations}
within $10^{-3}\%$ interpolation error.

\begin{figure}[H]\begin{center}
\includegraphics[scale=0.65]{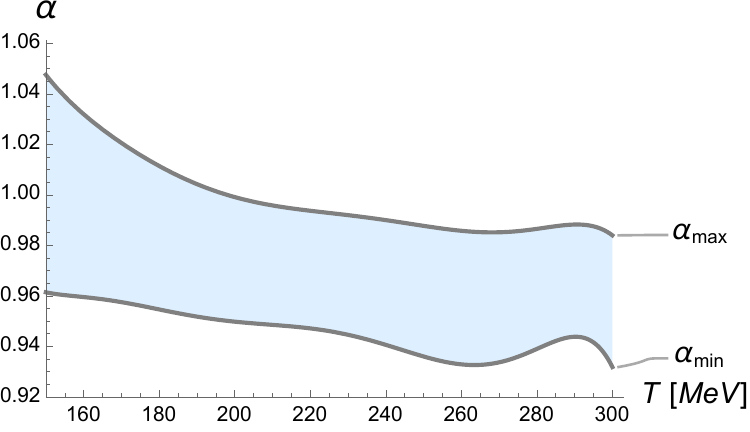}
\caption{\footnotesize Range of the parameter $\alpha$ in the deformed AdS$_5$-Schwarzschild black brane \eqref{1}, with coefficients (\ref{eq:Nu}, \ref{eq:Au}), as a function of the QGP temperature (MeV), using the experimental results of the QGP at LHC by the Jyv\"askyl\"a-Helsinki-Munich group \cite{Parkkila:2021yha} for $\eta/s$.}
\label{gmm33}\end{center}
\end{figure}

 Now, the up-to-date experimental results of the shear viscosity-to-entropy density ratio of the QGP, analyzed by the  Jyv\"askyl\"a-Helsinki-Munich group \cite{Parkkila:2021yha}, can also be used to constrain deformations of the AdS$_5$-Schwarzschild black brane. 
This scrutiny addressed transport coefficients of the QGP created in relativistic heavy-ion collisions, unraveled by an improved global Bayesian using the LHC Pb-Pb data at 5.02 TeV \cite{ALICE:2016ccg,ALICE:2019hno}. The uncertainty of the shear viscosity significantly decreases, by the inclusion of collective flow observables from two collision energies. Similar to the previous cases, matching the analytical form of $\eta/s$ \eqref{eq:etaSfinal} for the deformed AdS$_5$-Schwarzschild  black brane to the experimental values of $\eta/s$ of the QGP in Fig. 3 of Ref. \cite{Parkkila:2021yha} yields the minimum and maximum values of the deformation parameter $\alpha$, for each value of the temperature in the range $150\,{\rm MeV}\lesssim T \lesssim 300\,{\rm MeV}$. 
The bound for $\alpha$, as a function of the temperature, is shown in Fig. \ref{gmm33}. 
\noindent At the Hagedorn temperature, the range $0.963\lesssim \alpha\lesssim 1.047$ can be read off. 
We can interpolate the extremal values of the bound for $\alpha$ in Fig. \ref{gmm33}, as:
\begin{subequations}
\beq
\alpha_{\textsc{min}}(T)&=&-6.9706\times 10^{-15} T^7+1.0449\times 10^{-11} T^6-6.64188\times 10^{-9}
   T^5+2.32077\times 10^{-6} T^4\nonumber\\
   &&\qquad\,4.81445\times 10^{-4}
   T^3+5.93037\times 10^{-2} T^2-4.01706 T+116.431,\\
  \alpha_{\textsc{max}}(T)&=&-3.77795\times 10^{-15} T^7+5.77433\times 10^{-12} T^6-3.74585\times 10^{-9}
   T^5+1.33709\times 10^{-6} T^4\nonumber\\
   &&\qquad\,-2.83736 \times 10^{-4} 
   T^3+3.58205\times 10^{-2}  T^2-2.49468 T+75.1252,
   \eeq\end{subequations}
within $10^{-3}\%$ interpolation error. 

\begin{figure}[H]\begin{center}
\includegraphics[scale=0.65]{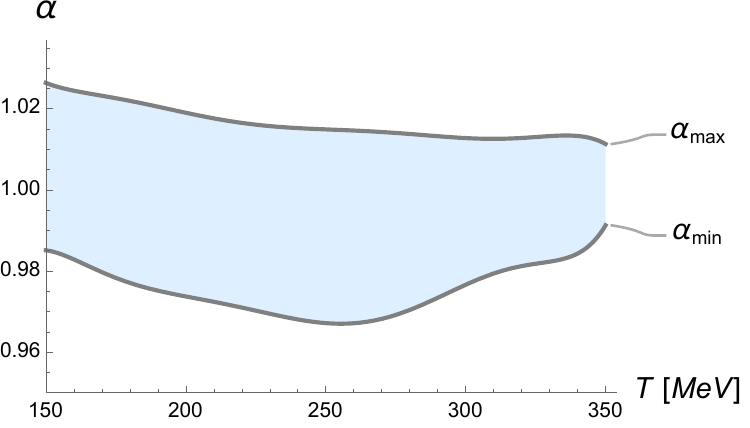}
\caption{\footnotesize \footnotesize Range of the parameter $\alpha$ in the deformed AdS$_5$-Schwarzschild black brane \eqref{1}, with coefficients (\ref{eq:Nu}, \ref{eq:Au}), as a function of the QGP temperature (MeV),  using the analysis of experimental data of the QGP at LHC by the MIT-Utrecht-Gen\`eve group using the \textsc{Trajectum}  framework for $\eta/s$.}
\label{gmmtra}\end{center}
\end{figure}

Finally, the analysis of experimental data of the QGP at LHC by the MIT-Utrecht-Gen\`eve group using the \textsc{Trajectum}  framework, with an improved global Bayesian analysis of the LHC Pb-Pb data at $\sqrt{s_{NN}} =  2.76$ and 5.02 TeV can be employed to bound deformations of the AdS$_5$-Schwarzschild black brane. The analysis in Ref. \cite{Nijs:2022rme} showed a stronger temperature dependence of the QGP shear viscosity, taking into account the latest measurements of higher harmonics in the hydrodynamical flow as well as the higher-order flow fluctuation observables, acting as inputs in the Bayesian analysis. 
The range of the temperature-dependent $\eta/s$ ratio in the left bottom panel in Fig. 3 of Ref. \cite{Nijs:2022rme} can be compared to the shear viscosity-to-entropy density ratio of the deformed AdS$_5$-Schwarzschild black brane, in Eq. \eqref{eq:etaSfinal}, and solved to obtain the minimum and maximum values of the parameter $\alpha$. The temperature-dependent bound on $\alpha$ is plotted in Fig. \ref{gmmtra}.
\noindent 
At the Hagedorn temperature, analysis of the experimental value of $\eta/s$ yields the range $0.984\lesssim \alpha\lesssim 1.026$ for the parameter regulating deformations of the AdS$_5$-Schwarzschild black brane. 
The temperature-dependent maxima and minima of $\alpha$ in Fig. \ref{gmmtra} can be interpolated by the expressions
\begin{subequations}
\beq\label{mit1}
\alpha_{\textsc{min}}(T)&=&6.57587\times 10^{-16} T^7-1.11752\times 10^{-12} T^6+8.01756\times 10^{-10}
   T^5-3.14694\times 10^{-7} T^4\nonumber\\
   &&\qquad\,+7.29766\times 10^{-5}
   T^3-9.99758\times 10^{-3} T^2+0.74893T-23.6695,\\
     \alpha_{\textsc{max}}(T)&=&-2.67183\times 10^{-16} T^7+4.58546\times 10^{-13} T^6-3.32588\times 10^{-10}
   T^5+1.32076\times 10^{-7} T^4\nonumber\\
   &&\qquad\,3.09995 \times 10^{-5} 
   T^3+4.29931\times 10^{-3}  T^2-0.32639T+11.4997,\label{mit2}
   \eeq\end{subequations}
within $10^{-3}\%$ interpolation error. 

Taking into account the range of values consistent with $\eta/s>0$, near the Hagedorn temperature the most severe range for the parameter driving deformations of the AdS$_5$-Schwarzschild black brane is given by the intersection of all allowed ranges in Figs. \ref{gmm22} -- \ref{gmmtra}, consisting of
\beq\label{1011}
1\leq \alpha \lesssim 1.0143.
\eeq Since the standard AdS$_5$-Schwarzschild black brane corresponds to the limit $\alpha \to 1$ of the deformed AdS$_5$-Schwarzschild black brane \eqref{1}, with coefficients (\ref{eq:Nu}, \ref{eq:Au}), one can conclude that 
the analysis of experimental values of $\eta/s$ of the QGP yields a very narrow range for deformations of the AdS$_5$-Schwarzschild black brane. The standard AdS$_5$-Schwarzschild black brane is thus robust against deformations, however, there is still a margin for mild deformations to analyze the QGP in the context of the membrane paradigm of AdS/CFT.   

Also, in the mimimal value of the range \eqref{1011}, corresponding to the limit $\alpha \to1$, the deformed AdS$_5$-Schwarzschild  black brane shear viscosity-to-entropy ratio is precisely the KSS viscosity-bound saturation ${\eta}/{s}={1}/{4\pi}\approxeq7.95774\times 10^{-2}$.
On the other hand, the maximal value $\alpha \to1.014$ of the range \eqref{1011} makes  ${\eta}/{s}={1}/{4\pi}\approxeq6.83259\times 10^{-2}$.
It means that the analysis of the experimental data of the QGP yields ${\eta}/{s}$ to attain values at most 14.14\% lower than the  KSS viscosity-bound saturation.

\section{Concluding remarks and perspectives} \label{sec:5}

A family of deformed AdS$_5$-Schwarzschild black brane was obtained from the embedding procedure, where the effective Einstein's equations in the vacuum are corrected by the electric part of the Weyl tensor, encoding a Weyl-type fluid. 
The deformed black brane associated thermodynamical functions were obtained from the bulk Einstein--Hilbert action with negative cosmological constant, with a Gibbons--Hawking term and a counterterm that prevents divergences.
Both the deformed black brane Hawking temperature and the existence of a real event horizon constrained the deformation parameter $\alpha$. Besides, 
an off-diagonal perturbation and the analysis of the 
retarded two-point function for the boundary CFT energy-momentum tensor provided a mechanism to derive the shear viscosity-to-entropy density ratio of the deformed AdS$_5$-Schwarzschild black brane metric. With the additional constraint that the event horizon is of Killing type, together with the Hagedorn temperature obtained from both the AdS/QCD hard- and soft-wall models, the allowed range for the deformation parameter $\alpha$ took a stricter form. The temperature-dependent shear mode damping constant was also computed and discussed.  The deformation parameter was severely more constrained by the ulterior analysis of experimental data of the QGP  
at LHC and RHIC. We took the calculation of $\eta/s$ accomplished by the JETSCAPE Bayesian model  \cite{JETSCAPE:2020shq},  the results by the Duke group \cite{Bernhard:2019bmu}, the calculations by the  Jyv\"askyl\"a-Helsinki-Munich group \cite{Parkkila:2021yha}, and also the results by the MIT-Utrecht-Gen\`eve group \cite{Nijs:2022rme,Nijs:2022rme}. Comparing 
the results by these groups and to the $\eta/s$ ratio of deformed AdS$_5$-Schwarzschild black brane metric
in Eq. \eqref{eq:etaSfinal}, the temperature-dependent minima and maxima of the deformation parameter $\alpha$ were calculated and discussed, whose most severe range 
is given by the inequality \eqref{1011}. With it, we showed that the family of deformed AdS$_5$-Schwarzschild black branes is robust against deformations, but a narrow range of deformations is still permitted. It may probe additional properties of QCD matter in the AdS/QCD approach.

The family of deformed AdS$_5$-Schwarzschild black branes, here obtained by the embedding procedure, violates the KSS viscosity bound in the range (\ref{1011}),  wherein the  parameter regulating the family of deformed AdS$_5$-Schwarzschild black branes complies with experimental data of the $\eta/s$ for the QGP  
at LHC and RHIC. 
The violation of the KSS saturation bound by the family of deformed AdS$_5$-Schwarzschild black branes comes from the fact that the metric \eqref{1}, with coefficients (\ref{eq:Nu}, \ref{eq:Au}), arises from the embedding protocol of the AdS$_5$ spacetime, as exhibited in the Appendix \ref{app2}. Appendix \ref{app3} presents an alternative setup, generating the family of deformed AdS$_5$-Schwarzschild black branes as exact solutions of a gravitational action beyond GR, suggesting 2-loop quantum corrections in the dual gravity sector. From this point of view, the dual viscous hydrodynamics carry such effects through the thermodynamic state functions and transport coefficients.

\section*{Acknowledgements}
RdR~thanks to The S\~ao Paulo Research Foundation -- FAPESP
(Grants No. 2021/01089-1 and No. 2024/05676-7) and to the National Council for Scientific and Technological Development -- CNPq  (Grants No. 303742/2023-2 and No. 401567/2023-0), for partial financial support.

\appendix

\section{The deformed AdS$_{5}$-Schwarzschild black brane}
\label{app2}

\par A correspondence between AdS/CFT and braneworld scenarios can be established in the membrane paradigm. A bottom-up approach for obtaining the deformed AdS$_{5}$-Schwarzschild black brane consists of embedding the AdS$_5$ apparatus into an AdS$_6$ bulk. 
One denotes by   $\gamma_{\scalebox{.6}{$AB$}}$ the AdS$_{6}$ metric and by  $g_{\scalebox{.6}{$AB$}}$ the metric equipping AdS$_{5}$, induced by   $\gamma_{\scalebox{.6}{$AB$}}$ as
\beq
g_{\scalebox{.6}{$AB$}} = \gamma_{\scalebox{.6}{$AB$}} + n_{\scalebox{.6}{$A$}}n_{\scalebox{.6}{$B$}},
\eeq
implementing the projection onto AdS$_{5}$.
The extrinsic curvature is defined as the Lie derivative of the metric, as $K_{\scalebox{.6}{$AB$}} = \frac{1}{2}\mathcal{L}_n  g_{\scalebox{.6}{$AB$}}=\frac12 n^{\scalebox{.6}{$C$}} \nabla_{\scalebox{.6}{$C$}} g_{\scalebox{.6}{$AB$}}$, which can be recast as
\beq\label{extc}
K_{\scalebox{.6}{$AB$}}  &=& -g_{\scalebox{.6}{$A$}}^{\;\scalebox{.6}{$C$}}g_{\scalebox{.6}{$B$}}^{\;\scalebox{.6}{$D$}}\nabla_{\scalebox{.6}{$C$}} n_{\scalebox{.6}{$D$}}.\nonumber
\eeq
\begin{figure}[H]
\centering\includegraphics[width=12.1cm]{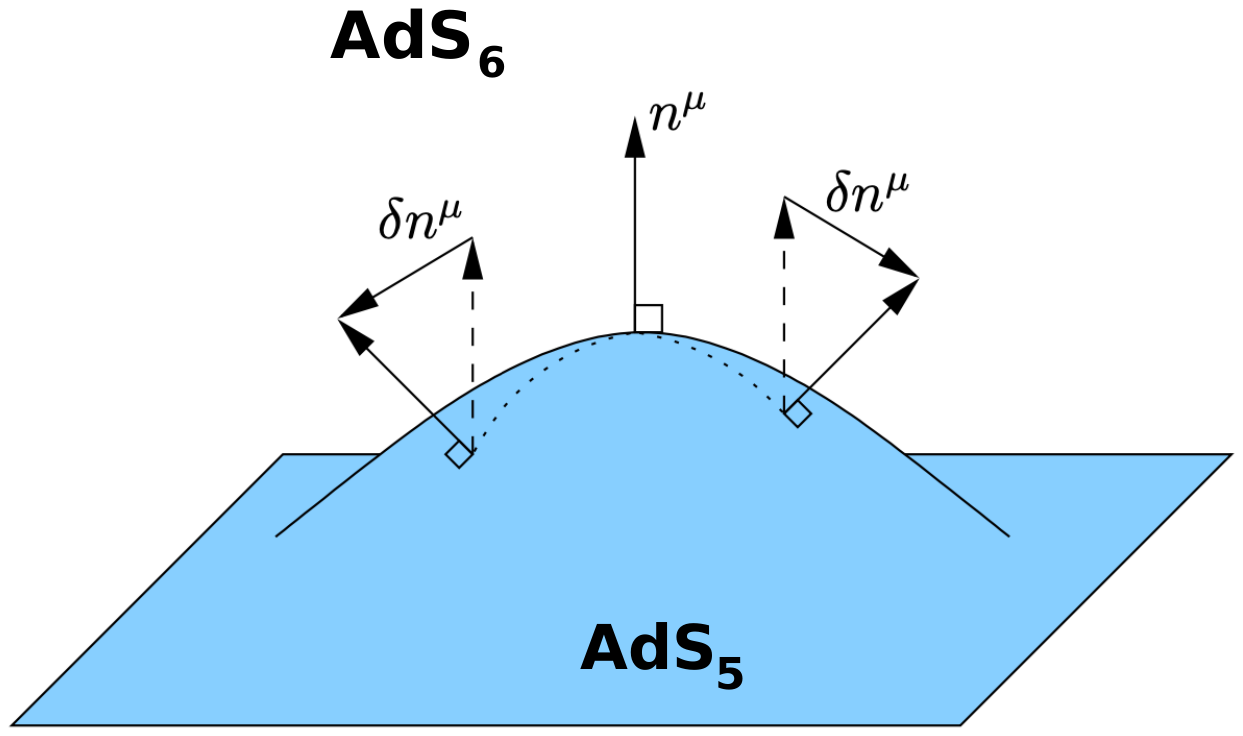}\vspace*{-1cm}
\caption{\footnotesize{Embedding of AdS$_{5}$ in AdS$_{6}$.}}
\end{figure} 
\noindent The AdS$_6$ bulk cosmological constant $\Uplambda$ enters  the effective Einstein's equations 
\begin{eqnarray}
R_{\scalebox{.6}{$MN$}}=\Uplambda\,g_{\scalebox{.6}{$MN$}}+\mathcal{E}_{\scalebox{.6}{$MN$}}
\,,
\label{D+1eq}
\end{eqnarray}
where ${\scalebox{.78}{$M, N$}}= 0,1,2,3,5,6$. The  AdS$_5$ manifold has a Gaussian 
coordinate chart $x^{\scalebox{.6}{$A$}}$ = ($x^\mu, x^5$), for $x^5=r$ denoting the coordinate along AdS$_5$, out of the AdS$_5$ boundary, which lives at $r=0$.
Einstein's equations on AdS$_5$ carry high-energy and non-local corrections 
from the embedding onto AdS$_6$, encoded by  
an AdS Weyl fluid flow  \cite{ssm1}. This hydrodynamical fluid flow can be implemented by the AdS$_6$ Weyl tensor, whose projection onto AdS$_5$ yields the electric part of the Weyl tensor, expressed as\footnote{The parameter $\kappa$ hereon denotes the tension of the black brane, seen as a membrane with surface viscosity in the membrane paradigm
\cite{Iqbal:2008by,Parikh:1997ma,Price:1986yy,hub}. 
}
\begin{eqnarray}
\!\!\!\!\!\!\!\!\mathcal{E}_{\scalebox{.6}{$AB$}}(\kappa^{-1}) \!&=&\!-\frac{6}{\kappa}\!\left[ \mathcal{U}\!\left(\!u_{\scalebox{.6}{$A$}} u_{\scalebox{.6}{$B$}} \!+\! \frac{1}{3}h_{\scalebox{.6}{$AB$}}\!\right)+ \mathit{Q}_{({\scalebox{.6}{$A$}}} u_{{\scalebox{.6}{$B$}})}\!+\!\mathcal{P}_{\scalebox{.6}{$AB$}}\right], \label{A4}
\end{eqnarray}
\noindent for ${\scalebox{.88}{$A, B$}} =0,1,2,3,5$. In Eq. (\ref{A4}), the term \beq\mathcal{U}=-\frac16\kappa\mathcal{E}_{\scalebox{.6}{$AB$}} u^{\scalebox{.6}{$A$}} u^{\scalebox{.6}{$B$}}\eeq denotes the effective energy density mimicking dark radiation, and
\beq
\mathcal{P}_{\scalebox{.6}{$AB$}}=-\frac16\kappa\left(h_{({\scalebox{.6}{$A$}}}^{\;{\scalebox{.6}{$C$}}}h_{{\scalebox{.6}{$B$}})}^{\;{\scalebox{.6}{$D$}}}-\frac13 h^{\scalebox{.6}{$CD$}}h_{\scalebox{.6}{$AB$}}\right)\mathcal{E}_{\scalebox{.6}{$CD$}}\eeq is the non-local anisotropic stress-tensor, whereas the non-local energy flux reads 
\beq\mathit{Q}_{\scalebox{.6}{$A$}} = -\frac16\kappa p^{\;{\scalebox{.6}{$B$}}}_{{\scalebox{.6}{$A$}}}\mathcal{E}_{\scalebox{.6}{$BC$}}u^{\scalebox{.6}{$C$}}.\eeq 
The electric part of the Weyl tensor can be alternatively expressed by \cite{,Meert:2020sqv}
\beq
\mathcal{E}_{\scalebox{.6}{$AB$}}=-\frac{\Uplambda_{5}}{6}g_{\scalebox{.6}{$AB$}}-\partial_{x^6} K_{\scalebox{.6}{$AB$}}+K_{\scalebox{.6}{$A$}}^{\scalebox{.6}{$\;\,C$}}K_{\scalebox{.6}{$CB$}},
\eeq
where $x^6$ denotes the Gaussian coordinate along the AdS$_6$ bulk.

  Denoting by $G_{\scalebox{.6}{$AB$}}$ the Einstein tensor, the vacuum 5-dimensional Einstein's equations acquire a correction due to the Weyl fluid, witting 
\begin{equation}
G_{\scalebox{.6}{$AB$}} + \Lambda_5 g_{\scalebox{.6}{$AB$}}
=\mathcal{E}_{\scalebox{.6}{$AB$}}.
\label{projeinstein}
\end{equation} 
Since $\mathcal{E}_{\scalebox{.6}{$AB$}} \sim \kappa^{-1}$, GR is recovered in AdS$_5$,  in the limit  $\kappa \rightarrow \infty$, and the Einstein's field equations have the standard form  $G_{\scalebox{.6}{$AB$}} + \Lambda_5 g_{\scalebox{.6}{$AB$}}= 0$, with the standard AdS$_5$-Schwarzschild black brane as a solution. 
In the decoupling limit $8\pi G_5 g_{\scalebox{.65}{\textsc{YM}}}^2L^2\gg 1$ the deformed AdS$_{5}$-Schwarzschild black brane can be considered as a fixed background, with any backreaction suppressed. 

On the one hand, one can work with a system of equations that is a weaker requirement than vacuum Einstein's equations in AdS$_5$: 
\begin{eqnarray}
\tilde{R}_{{\scalebox{.62}{$A$}}6}=0,\qquad\qquad
{R}=\Lambda_5,
\label{Deq}
\end{eqnarray} where the Ricci tensor in AdS$_6$ is denoted by $\tilde{R}_{{\scalebox{.6}{$MN$}}}$. 
Eqs.~(\ref{Deq}) also emulate constraints in the ADM method and the field equations $R_{\scalebox{.6}{$AB$}}=\mathcal{E}_{\scalebox{.6}{$AB$}}$ close the system. The ADM procedure takes into account the Gauss equation, which relates the Riemann tensor of  AdS$_6$ to the sum of the Riemann tensor of  AdS$_5$ and a quadratic antisymmetrized combination of the extrinsic curvature, as
   \begin{eqnarray}
{}^{(6)}R^{\scalebox{.6}{$A$}} _{\;\;\scalebox{.6}{$BCD$}} = {}^{(5)}R^{\scalebox{.6}{$A$}}_{\;\;\scalebox{.6}{$BCD$}}
 -K^{\scalebox{.6}{$A$}} _ {\;\; {\scalebox{.6}{$C$}}}K_{{\scalebox{.6}{$BD$}}} + K^{\scalebox{.6}{$A$}} _ {\;\; {\scalebox{.6}{$D$}}}K_{{\scalebox{.6}{$BC$}}},\label{gauss}\end{eqnarray} 
which can be contracted with the induced metric in   AdS$_{5}$. Using Einstein's equations, one arrives at the  Hamiltonian constraint
   \begin{eqnarray}\label{eul}
\mathcal{H}\equiv {}^{(5)}R + K^2 - K_{\scalebox{.6}{$AB$}}K^{\scalebox{.6}{$AB$}} - 16\pi n^{\scalebox{.6}{$A$}} n^{\scalebox{.6}{$B$}} T_{\scalebox{.6}{$AB$}} = 0.\label{gaussh}\end{eqnarray} 
The last term on the left-hand side of Eq. (\ref{eul}) encodes the  total energy density measured by an Eulerian observer. 

On the other hand, the Codazzi equations are given by  
   \begin{eqnarray}
{}^{(6)}R_{{\scalebox{.6}{$ABCD$}}}n^{\scalebox{.6}{$D$}} = 
 \nabla_{\scalebox{.6}{$B$}} K_{\scalebox{.6}{$AC$}} - \nabla_{\scalebox{.6}{$A$}} K_{{\scalebox{.6}{$BC$}}},\label{gausscodazzi}\end{eqnarray} 
which can be contracted with the induced metric of AdS$_{5}$, yielding the momentum constraint:
   \begin{eqnarray}\label{eul1}
\mathcal{M}^{\scalebox{.6}{$A$}}\equiv \nabla_{\scalebox{.6}{$B$}} K^{\scalebox{.6}{$AB$}} - D^{\scalebox{.6}{$A$}} K - 8\pi g^{\scalebox{.6}{$AB$}} n^{\scalebox{.6}{$C$}}  T_{{\scalebox{.6}{$BC$}}} = 0.\label{gauss1}\nonumber\end{eqnarray} 
The last term on the left-hand side of Eq. (\ref{eul1}) represents the momentum density measured by an Eulerian observer. 

Summing the momentum and the Hamiltonian constraints
yields Eq. (\ref{ctr}), whose right-hand side has the following expression:
{{\begin{eqnarray}\label{r1000}
k(r, r_{\scalebox{.65}{$0$}}, \alpha)\!&\!=\!&\!-\frac{1}{r^{10}}\Bigg\{-\left[10 (\alpha -1)+r^6-3 r^2r_{\scalebox{.65}{$0$}}^4\right] \left(\alpha +r^6-r^2
   r_{\scalebox{.65}{$0$}}^4-1\right)+\frac{4 r^8 \left(-2 \alpha +r^6+r^2
   r_{\scalebox{.65}{$0$}}^4+2\right)^2}{\left(\alpha +r^6-r^2 r_{\scalebox{.65}{$0$}}^4-1\right)^2}\nonumber\\
  && +\frac{4 r^8 \left[4 r^{12}\!+\!8 (2\!-\!3 \alpha ) r^8 r_{\scalebox{.65}{$0$}}^4\!+\!(20 \alpha \!-\!23) r^4
   r_{\scalebox{.65}{$0$}}^8\!+\!3 (4 \alpha \!-\!1) r_{\scalebox{.65}{$0$}}^{12}\right]^2}{\left(2 r^8-5 r^4 r_{\scalebox{.65}{$0$}}^4+3 r_{\scalebox{.65}{$0$}}^8\right)^2 \left(2
   r^4+(1-4 \alpha ) r_{\scalebox{.65}{$0$}}^4\right)^2}\nonumber\\
  && -\frac{2 r^8 \left[8 r^{16}-60
   r^{12} r_{\scalebox{.65}{$0$}}^4+6 (40 \alpha  (2 \alpha -3)+67) r^8 r_{\scalebox{.65}{$0$}}^8+(4 \alpha -1) (20 \alpha +43) r^4
   r_{\scalebox{.65}{$0$}}^{12}-9 (1-4 \alpha )^2 r_{\scalebox{.65}{$0$}}^{16}\right]}{\left(2 r^8-5 r^4 r_{\scalebox{.65}{$0$}}^4+3 r_{\scalebox{.65}{$0$}}^8\right) \left(2
   r^4+(1-4 \alpha ) r_{\scalebox{.65}{$0$}}^4\right)^2}\nonumber\\&&+\frac{1}{2
   r^4\!+\!(1\!-\!4 \alpha ) r_{\scalebox{.65}{$0$}}^4}\left[r^2 \left(2 r^8\!+\!2 r^6\!-\!5 r^4
   r_{\scalebox{.65}{$0$}}^4\!+\!(1\!-\!4 \alpha ) r^2 r_{\scalebox{.65}{$0$}}^4\!+\!3 r_{\scalebox{.65}{$0$}}^8\right) \left(\alpha \!+\!r^6\!-\!r^4\!-\!r^2 r_{\scalebox{.65}{$0$}}^4\!-\!1\right)\right]\nonumber\\&&+\frac{4 r^8 \left(r^6\!+\!r^2 r_{\scalebox{.65}{$0$}}^4\!+\!2\!-\!2\alpha\right) \left(4
   r^{12}\!+\!8 (2\!-\!3 \alpha ) r^8 r_{\scalebox{.65}{$0$}}^4\!+\!3 (4 \alpha \!-\!1)
   r_{\scalebox{.65}{$0$}}^{12}\right)}{\left(2 r^4\!-\!3 r_{\scalebox{.65}{$0$}}^4\right) \left(r^4\!-\!r_{\scalebox{.65}{$0$}}^4\right) \left(2 r^4\!+\!(1\!-\!4 \alpha )
   r_{\scalebox{.65}{$0$}}^4\right) \left(\alpha\!+\!r^6\!-\!r^2 r_{\scalebox{.65}{$0$}}^4\!-\!1\right)}\nonumber\\
  &&  
   +2 r^8 \left(\frac{2 r^8+5 r^4 r_{\scalebox{.65}{$0$}}^4-9 r_{\scalebox{.65}{$0$}}^8}{2 r^8-5 r^4 r_{\scalebox{.65}{$0$}}^4+3 r_{\scalebox{.65}{$0$}}^8}-\frac{4 r^4}{2 r^4+(1-4 \alpha )
   r_{\scalebox{.65}{$0$}}^4}+\frac{r^2 \left(3
   r^4-r_{\scalebox{.65}{$0$}}^4\right)}{\alpha +r^6-r^2 r_{\scalebox{.65}{$0$}}^4-1}\right)\Bigg\}.
   \end{eqnarray}}}

\section{deformed AdS$_5$-Schwarzschild  black brane as an exact solution of higher-order curvature terms}\label{app3}

A thorough examination of the numerical values allowed for the deformation $\alpha$, culminating in the interval (\ref{1011}) obtained from the analysis of the QGP experimental data,  shows that the $\eta/s$ ratio \eqref{eq:etaSfinal} of the deformed AdS$_5$-Schwarzschild  black brane does violate the KSS viscosity bound. 
This violation emerged within the embedding protocol, where the sum of the momentum and the Hamiltonian constraints (\ref{gauss}, \ref{gauss1}) imply the EDO (\ref{ctr}), whose solution is the deformed AdS$_5$-Schwarzschild  black brane metric (\ref{1}), with coefficients \eqref{eq:Nu} and \eqref{eq:Au}. 

Alternatively, 
the KSS viscosity-bound violation complies with a gravitational background beyond the Einstein--Hilbert action. 
In fact, invariance under diffeomorphism comprises the fundamental symmetry of any action that describes gravity.
In this context, the Einstein--Hilbert action represents the lowest-energy, deep infrared, approximation of a more general action involving higher-order curvature operators that can be constructed upon the Riemann tensor, the Ricci tensor, and the scalar curvature. 
The quantum fluctuations of a field in a curved spacetime give rise to UV divergences that take the form of invariants of the metric and curvature in the quantum effective action. 
Any higher-order operators beyond the Einstein--Hilbert term are led up by dimensionful coefficients. Therefore, these operators are intangible in the context of the renormalization group. That way, one can explore the fate of the deformed AdS$_5$-Schwarzschild  black branes, seen as solutions to the equations of motion associated with a more general action involving higher-order curvature operators, 
corresponding to quantum corrections to GR.
Higher-curvature, or more widely, higher-derivative
interactions are expected to emerge  as stringy or quantum 
corrections to the classical action describing gravity. Consequently, a more refined prescription can be given by an
effective field theory action for gravity where the Einstein--Hilbert term with negative cosmological constant is  supplemented by higher-curvature corrections. This effective action may yield a reliable dual superconformal gauge theory. Effective 5-dimensional gravity is described by a sensible derivative expansion, in a way that powers of the Planck length systematically suppress the higher-curvature terms. 
The potentially infinite sum of operators must be truncated somehow, for the task of deriving the equations of motion to be feasible and physically sound.
The 1-loop renormalization of Einstein--Hilbert gravity was implemented by ’t Hooft and Veltman \cite{tHooft:1974toh} and the 2-loop one was proposed by
Goroff and Sagnotti \cite{Goroff:1985th}. 
Refs. \cite{Lessa:2023thi,Bueno:2016lrh} studied Einsteinian cubic gravity, whereas Ref. \cite{Shapiro:2015uxa} explored Lee--Wick gravity. We do not aim here to explore renormalizability but solely address the family of deformed AdS$_5$-Schwarzschild  black branes as formal solutions of higher-curvature terms in the gravitational action. 
%
 
For any isotropic holographic model with an effective gravitational action with at most two derivatives of the metric, the shear viscosity-to-entropy density ratio satisfies $\eta/s \gtrsim  1/4\pi$ \cite{kss}. Sufficiently generic corrections to the Einstein--Hilbert action might violate the KSS viscosity bound \cite{Schafer:2009dj}. Now, since corrections to Einstein's gravity are expected in a quantum theory of gravity, the KSS viscosity-bound violation emulates a constraint on possible corrections to Einstein's gravity. 
Ref. \cite{viol1} studied an action for gravity with Gauss--Bonnet term.  To avoid microcausality violation, 4-dimensional boundary gauge theories dual to 5-dimensional Gauss--Bonnet gravity have ${\eta}/{s}$  $\sim36\%$ smaller than the original KSS viscosity bound, representing a violation to the conjectured universal bound. Besides, scalar-Gauss--Bonnet gravity also predicts $\eta/s<1/4\pi$. Also, the violation of the KSS viscosity bound was implemented in Ref. \cite{Bravo-Gaete:2022lno} in the context of degenerate higher-order scalar-tensor theories \cite{DeFelice:2018ewo,Feng:2015oea}. In the membrane paradigm, the shear viscosity, transverse to the direction of anisotropy, saturates the KSS viscosity bound. The longitudinal shear viscosities are smaller, providing an example that does not involve higher-derivative theories of gravity, with gauge/gravity correspondence fully known \cite{Rebhan:2011vd}. Ref. \cite{Kats:2007mq} scrutinized the effects of curvature squared corrections in AdS on the shear viscosity of the dual gauge field theory. Ref. \cite{Banerjee:2009wg} considered in this context general four derivative terms and quartic Weyl terms arising in type IIB string theory, showing that the shear viscosity of the fluid on the boundary corresponds to the near-horizon value of the transverse graviton effective coupling when the action has arbitrary
higher-derivative terms in the bulk.

For the 
Ricci and Einstein cubic gravity to encompass terms that are not topological, alternatively, to the embedding protocol, the generalized extremal brane described by the metric \eqref{1}, with coefficients (\ref{eq:Nu}, \ref{eq:Au}), is an exact solution of the equations of motion coming from the action 
\beq
S_{\tiny{(3)}} = \int {\scalebox{.95}{$\mathrm{d}$}}^5x \sqrt{-g}\;{H_3}\left(R,R_{\scalebox{.6}{$AB$}},R_{\scalebox{.6}{$ABCD$}}\right)+\lim_{u\to 0}\int \!{\scalebox{.95}{$\mathrm{d}$}}^{4}x\sqrt{-g}K+S_{\textsc{c.t}}\,,\label{cubic1}
\eeq
where
\beq
 {H_3} &=& \left(R-2\Lambda_5\right)+\alpha_1G_{\scalebox{.6}{$AB$}}\Box R^{\scalebox{.6}{$AB$}}  +\alpha_2\left(-\frac{65}{324}R^3+\frac{29}{27}RR_{\scalebox{.6}{$AB$}}R^{\scalebox{.6}{$AB$}}-\frac{59}{81}R_{\scalebox{.6}{$A$}}^{\scalebox{.6}{$\;B$}}R_{{\scalebox{.6}{$C$}}}^{\,{\scalebox{.6}{$A$}}}R_{{\scalebox{.6}{$B$}}}^{\;{\scalebox{.6}{$C$}}}+14R_{{\scalebox{.6}{$AB$}}}^{\;\,{\scalebox{.6}{$CD$}}}R_{{\scalebox{.6}{$CD$}}}^{\;\,{\scalebox{.6}{$EF$}}}R_{{\scalebox{.6}{$EF$}}}^{\;\,{\scalebox{.6}{$CD$}}}\right.\nonumber\\
&&\left.
\qquad\qquad\qquad\qquad\qquad\qquad\qquad-4R_{{\scalebox{.6}{$ABCD$}}}R^{{\scalebox{.6}{$ABC$}}}_{{\scalebox{.6}{$E$}}}R^{{\scalebox{.6}{$DE$}}}
-\frac7{108}R_{{\scalebox{.6}{$ABCD$}}}R^{{\scalebox{.6}{$ABCD$}}}R+4R_{{\scalebox{.6}{$ABCD$}}}R^{{\scalebox{.6}{$AC$}}}R^{{\scalebox{.6}{$BD$}}}\right)\nonumber\\
&&
\qquad\qquad+\alpha_3\left(\frac45\nabla_{\scalebox{.6}{$A$}} R_{{\scalebox{.6}{$BC$}}}\nabla^{\scalebox{.6}{$A$}} R^{{\scalebox{.6}{$BC$}}}+\frac7{18}\nabla_{\scalebox{.6}{$A$}} R_{{\scalebox{.6}{$BC$}}}\nabla^{\scalebox{.6}{$C$}} R^{{\scalebox{.6}{$AB$}}}+\frac{4}{9} \nabla_{\scalebox{.6}{$A$}} R \nabla^{\scalebox{.6}{$A$}} R+\frac{3}{2}\nabla_{\scalebox{.6}{$A$}} R_{{\scalebox{.6}{$BCDE$}}}\nabla^{\scalebox{.6}{$B$}} R^{{\scalebox{.6}{$ACDE$}}}\right.\nonumber\\
&&\left. \qquad\qquad\qquad\qquad - R^{{\scalebox{.6}{$AB$}}}\Box R_{{\scalebox{.6}{$AB$}}}+\frac3{8}R_{{\scalebox{.6}{$AB$}}}\nabla^{\scalebox{.6}{$A$}} \nabla^{\scalebox{.6}{$B$}} R+\frac7{18}R\nabla^{\scalebox{.6}{$A$}} \nabla^{\scalebox{.6}{$B$}} R_{{\scalebox{.6}{$AB$}}} \right),	\label{cubic}
\eeq 
\noindent 
with the other two terms in the action in Eq. (\ref{cubic1}) represent the Gibbons--Hawking action and the counterterm action  (\textsc{c.t.}). The fifth and sixth cubic terms in the parenthesis of the $\alpha_2$ coefficient were already discussed in Ref.  \cite{Buchel:2008vz}, which gave examples illustrating that the KSS viscosity bound is violated in superconformal gauge theories with non-equal central charges. Also, Ref. \cite{Bueno:2022log} proposed the terms in the last line of Eq. (\ref{cubic}), which naturally arise in a braneworld construction, corroborating the embedding protocol used here. 

Ref. \cite{Bueno:2016xff} stated that the rational coefficients accompanying each of the cubic terms in Eq. (\ref{cubic1}) make them neither topological nor trivial nor topological.  Ref. \cite{Modesto:2016ofr} demonstrated that the Lee--Wick term, whose coefficient is $\alpha_1$ in action (\ref{cubic1}, \ref{cubic}), is superrenormalizable, at least in four dimensions. An additional result that supports the existence of the deformed AdS$_5$-Schwarzschild black brane \eqref{1}, with coefficients (\ref{eq:Nu}, \ref{eq:Au}), is the exclusion of static and spherically symmetric black hole solutions,  with radial and temporal
 metric components being the negative inverse of each other, in Einsteinian cubic gravity with unsuppressed higher-order curvature terms \cite{DeFelice:2023vmj}.
About the terms in action (\ref{cubic1}, \ref{cubic}) accompanying the coefficient $\alpha_3$, their renormalizability should be explored, which is still an intricate task. This is not our aim here but to indicate the necessary steps to show that the deformed AdS$_5$-Schwarzschild black brane, as a solution to the action (\ref{cubic1}, \ref{cubic}), yields $\eta/s$ in Eq. \eqref{eq:etaSfinal} to be consistent with its experimental value dictated by data involving the QGP.  The associated QCD plasma is described by a dual effective conformal field theory, whose  holographic dual can be approximated by the action (\ref{cubic1}, \ref{cubic}) with controllable higher-curvature corrections.

Therefore, in this setup, the deformation parameter $\alpha$, which governs the deformed AdS$_5$-Schwarzschild  black brane metric (\ref{1}), with coefficients \eqref{eq:Nu} and \eqref{eq:Au},  is a rational combination of the coefficients $\alpha_1, \alpha_2$, and  $\alpha_3$ in Eq. (\ref{cubic}), respectively encoding quantum gravity/stringy effects beyond GR. To extend Ref. \cite{Goroff:1985th}  as an attempt to identify cubic gravity and the terms accompanying the coefficient $\alpha_3$ in action (\ref{cubic1}, \ref{cubic}) as 2-loop quantum corrections to 4-dimensional gravity, one must further investigate these terms.  We can assert that it shares the spectrum of Einstein's gravity, propagating a transverse and
massless graviton on a maximally symmetric background \cite{Bueno:2016xff}. Ref. \cite{Shapiro:2015uxa} has pointed out that the quantum effective action of weakly nonlocal theories likely has an
infinite number of complex conjugate poles. Therefore, one can focus on local higher-order curvature theories beyond Stelle’s theory, admitting the graviton and solely complex conjugate poles in the classical spectrum. These theories are unitary and in agreement with the Lee--Wick prescription. 
The QGP can be also studied in two holographic models for QCD based on either the Dirac--Born--Infeld brane prescription or cubic gravity-Maxwell-dilaton actions, extending Refs. \cite{Jokela:2024xgz,Fichet:2023dju,Silva:2023ieb}. 

Ref. \cite{Bravo-Gaete:2023iry} derived the most general formula for $\eta/s$ for gravitational actions beyond GR. From it, Eq. (\ref{eq:etaSfinal}) can be derived from the action (\ref{cubic1}, \ref{cubic}). 
The Kubo formula \cite{kss, Policastro:2001yc,Son:2002sd}
can be emulated for the action (\ref{cubic1}, \ref{cubic}), implementing metric perturbations in the bulk and obtaining the response function. One considers the first and second-order metric perturbations, respectively: 
\beq\delta^{(1)}g^{{\scalebox{.6}{$AB$}}}&=&-h^{{\scalebox{.6}{$AB$}}},\qquad\qquad\qquad
\delta^{(2)}g^{{\scalebox{.6}{$AB$}}}=h^{{\scalebox{.6}{$AC$}}}h^{{\scalebox{.6}{$B$}}}_{{\scalebox{.6}{$C$}}}\,.
\eeq 
The perturbations of the Ricci tensor at first order are given by
\begin{eqnarray}\label{1234}
\!\!\!\!\!\!\!\!\!\!\!\!\!\delta^{(1)}R_{{\scalebox{.6}{$AB$}}}&\!=\!&\partial_{{{\scalebox{.6}{$C$}}}}(\delta^{(1)}\Gamma^{{{\scalebox{.6}{$C$}}}}_{{\scalebox{.6}{$AB$}}})-\partial_{{{\scalebox{.6}{$A$}}}}(\delta^{(1)}\Gamma^{{{\scalebox{.6}{$C$}}}}_{{{\scalebox{.6}{$BC$}}}})+(\delta^{(1)}\Gamma^{{{\scalebox{.6}{$C$}}}}_{{{\scalebox{.6}{$CD$}}}})\Gamma^{{{\scalebox{.6}{$D$}}}}_{{\scalebox{.6}{$AB$}}}+\Gamma^{{{\scalebox{.6}{$C$}}}}_{{{\scalebox{.6}{$CD$}}}}(\delta^{(1)}\Gamma^{{{\scalebox{.6}{$D$}}}}_{{\scalebox{.6}{$AB$}}})\nonumber\\&&\qquad\qquad\qquad\qquad\qquad-(\delta^{(1)}\Gamma^{{{\scalebox{.6}{$C$}}}}_{{{\scalebox{.6}{$AD$}}}})\Gamma^{{{\scalebox{.6}{$D$}}}}_{{{\scalebox{.6}{$CB$}}}}-\Gamma^{{{\scalebox{.6}{$C$}}}}_{{{\scalebox{.6}{$AD$}}}}(\delta^{(1)}\Gamma^{{{\scalebox{.6}{$D$}}}}_{{{\scalebox{.6}{$CB$}}}}),
\eeq
where the perturbation of the Christoffel symbols in first and second order respectively read
\beq
\delta^{(1)}\Gamma^{{\scalebox{.6}{$C$}}}_{{\scalebox{.6}{$AB$}}}&=&\frac{1}{2}(\partial_{{\scalebox{.6}{$A$}}}h^{{\scalebox{.6}{$C$}}}_{{\scalebox{.6}{$B$}}}+\partial_{{\scalebox{.6}{$B$}}}h^{{\scalebox{.6}{$C$}}}_{{\scalebox{.6}{$A$}}}-\partial^{{\scalebox{.6}{$C$}}}h_{{\scalebox{.6}{$AB$}}}),\\
\delta^{(2)}\Gamma^{{\scalebox{.6}{$C$}}}_{{\scalebox{.6}{$AB$}}}&=&-\frac{1}{2}h^{{\scalebox{.6}{$CD$}}}(\partial_{{\scalebox{.6}{$A$}}}h_{{\scalebox{.6}{$BD$}}}+\partial_{{\scalebox{.6}{$B$}}}h_{{\scalebox{.6}{$AD$}}}-\partial_{{\scalebox{.6}{$D$}}}h_{{\scalebox{.6}{$AB$}}}).
\eeq
Therefore the perturbations of the Ricci tensor at second order wit
\beq
\delta^{(2)}R_{{\scalebox{.6}{$AB$}}}&=&\partial_{{{\scalebox{.6}{$C$}}}}(\delta^{(2)}\Gamma^{{{\scalebox{.6}{$C$}}}}_{{\scalebox{.6}{$AB$}}})-\partial_{{{\scalebox{.6}{$A$}}}}(\delta^{(2)}\Gamma^{{{\scalebox{.6}{$C$}}}}_{{{\scalebox{.6}{$C$}}}{{\scalebox{.6}{$B$}}}})+(\delta^{(1)}\Gamma^{{{\scalebox{.6}{$C$}}}}_{{{\scalebox{.6}{$CD$}}}})\delta^{(1)}\Gamma^{{{\scalebox{.6}{$D$}}}}_{{\scalebox{.6}{$AB$}}}-\delta^{(1)}\Gamma^{{{\scalebox{.6}{$C$}}}}_{{{\scalebox{.6}{$AD$}}}}\delta^{(1)}\Gamma^{{{\scalebox{.6}{$D$}}}}_{{{\scalebox{.6}{$CB$}}}}+\delta^{(2)}\Gamma^{{{\scalebox{.6}{$C$}}}}_{{{\scalebox{.6}{$CD$}}}}\Gamma^{{{\scalebox{.6}{$D$}}}}_{{\scalebox{.6}{$AB$}}}\nonumber\\
&&+\Gamma^{{{\scalebox{.6}{$C$}}}}_{{{\scalebox{.6}{$CD$}}}}\delta^{(2)}\Gamma^{{{\scalebox{.6}{$D$}}}}_{{\scalebox{.6}{$AB$}}}-\delta^{(2)}\Gamma^{{{\scalebox{.6}{$C$}}}}_{{{\scalebox{.6}{$AD$}}}}\Gamma^{{{\scalebox{.6}{$D$}}}}_{{{\scalebox{.6}{$CB$}}}}-\Gamma^{{{\scalebox{.6}{$C$}}}}_{{{\scalebox{.6}{$AD$}}}}\delta^{(2)}\Gamma^{{{\scalebox{.6}{$D$}}}}_{{{\scalebox{.6}{$CB$}}}}.
\eeq
Considering the first-order perturbations $\delta^{(1)}g_{{{\scalebox{.6}{$AB$}}}}=h_{{{\scalebox{.6}{$AB$}}}}$, one can express the transverse and traceless tensor perturbation to compute the shear viscosity, where $\partial_{{{\scalebox{.6}{$C$}}}}h_{{{\scalebox{.6}{$AB$}}}}=0$ and $h\equiv\eta^{{{\scalebox{.6}{$AB$}}}}h_{{{\scalebox{.6}{$AB$}}}}=0$. 
This  perturbation can be implemented in the 
deformed AdS$_5$-Schwarzschild black brane metric (\ref{1a}) and this setup can be used to investigate deformed AdS$_5$-Schwarzschild  black branes \eqref{1}, with coefficients (\ref{eq:Nu}, \ref{eq:Au}), and subsequently derive the ${\eta}/{s}$ ratio in this gravitational background. 
Consider an off-diagonal perturbation $\Psi\equiv h_{xy}$ of the metric (\ref{1}):
\begin{eqnarray}
\!\!\!\!\!\!\!\!\!d\mathsf{s}^2 = -\frac{r_{\scalebox{.65}{$0$}}^2}{u^2} F(u) {\scalebox{.95}{$\mathrm{d}$}}t^2 + \frac{1}{u^2 G(u)} {\scalebox{.95}{$\mathrm{d}$}}u^2 + \frac{r_{\scalebox{.65}{$0$}}^2}{u^2} \delta_{ij} {\scalebox{.95}{$\mathrm{d}$}}x^i {\scalebox{.95}{$\mathrm{d}$}}x^j +2h_{xy}{\scalebox{.95}{$\mathrm{d}$}}x{\scalebox{.95}{$\mathrm{d}$}}y. \label{1nt}
\end{eqnarray}
Eqs. (\ref{1234}, \ref{eq:Nu}, \ref{eq:Au}) yield
\begin{eqnarray}
\delta^{(1)}R_{xy}&=&-\frac{1}{2}\Box\Uppsi-\left[2\sqrt{-\frac{\left(3 u^8-5 u^4+2\right) \left((\alpha -1) u^6-u^4+1\right)}{(4 \alpha
   -1) u^4-2}}+\frac{H(u,\alpha)}{u}\right]\Uppsi,
\end{eqnarray}
where
\beq
H(u,\alpha)=\frac{J(u,\alpha)\,\left[(4 \alpha -1) u^4-2\right]}{\left((1-4 \alpha ) u^4+2\right)^2 \sqrt{-{\left(3 u^8-5
   u^4+2\right) \left((\alpha -1) u^6-u^4+1\right)}{}}},
\eeq
for 
\beq
J(u,\alpha) &\!=\!& 8 (\alpha +7) u^6-4 (\alpha +9) u^4+16 (\alpha -2) u^2+u^8 \left[-8 (\alpha -2)
   \alpha -6 (\alpha -1) (4 \alpha -1) u^{10}\right.\nonumber\\&&\left.-3 (4 \alpha -1) (5 \alpha -7) u^8+2 (5
   \alpha  (4 \alpha +1)-7) u^6+(\alpha  (60 \alpha -97)-23) u^4\right.\nonumber\\&&\left.-4 (4 \alpha  (\alpha
   +4)+1) u^2+66\right]\!+\!8.\eeq
 To compute $\eta$, the second order perturbation of the Einstein tensor $\delta^{(2)}G_{xy}$ will be employed, which is given by:
\begin{eqnarray}
\delta^{(2)}G_{xy}&=&-\frac{1}{8}\Box\Uppsi
+\frac{1}{8}\Uppsi\left[-\left[2\sqrt{-\frac{\left(3 u^8-5 u^4+2\right) \left((\alpha -1) u^6-u^4+1\right)}{(4 \alpha
   -1) u^4-2}}\right]+\frac{H(u,\alpha)}{u}\right]\Uppsi'\nonumber\\&&+\frac{1}{u}\sqrt{-\frac{\left(3 u^8-5 u^4+2\right) \left((\alpha -1) u^6-u^4+1\right)}{(4 \alpha
   -1) u^4-2}}\Uppsi''.
\end{eqnarray}
Combining all elements with Kubo’s formula relates the components of the viscosity to the two-point function of corresponding components of the stress-tensor $T_{xy}$, given by 
\begin{eqnarray}
\eta=-\frac{1}{\omega}\lim_{\substack{{\omega \rightarrow 0}\\{\mathfrak{q}_1}, {\mathfrak{q}_2} \to \bm{0}}}\Im\langle\,\left[T_{xy}(\mathfrak{q}_{1},\omega)T_{xy}(\mathfrak{q}_{2},\omega)\right]\,\rangle= -\frac{1}{\omega}\lim_{\substack{{\omega \rightarrow 0}\\{\mathfrak{q}_1}, {\mathfrak{q}_2} \to \bm{0}}}\Im\frac{\delta^{(2)}S_{(3)}}{\delta\,h_{xy}(\mathfrak{q}_{1})\delta\,h_{xy}(\mathfrak{q}_{2})},
\end{eqnarray}
for $S_{(3)}$ representing the action (\ref{cubic1}, \ref{cubic}). Finally, one can implement the receipt in Sec. 4 and Appendix C of Ref. \cite{Kats:2007mq} to verify that the shear viscosity-to-entropy ratio takes the form \eqref{eq:etaSfinal}. Therefore both the embedding protocol in Appendix \ref{app2} and the method presented in this section are consistent and yield the same result \eqref{eq:etaSfinal} for $\eta/s$.  
Determining the effect of higher-order curvature corrections
to thermal properties of the dual gauge theory is equivalent to studying the properties of deformed AdS$_5$-Schwarzschild  black brane within the action (\ref{cubic1}) with a universal set of higher-order curvature corrections. The results for $\eta/s$ will be universal for the class of theories described by the family of deformed AdS$_5$-Schwarzschild  black branes, emulating quantum corrections to $\eta/s$ in the dual field theory, as posed in Ref. \cite{Buchel:2008ae}.

\bibliography{bibliografia}
\bibliographystyle{iopart-num}

\end{document}